\documentclass[fleqn,usenatbib]{mnras}

\usepackage{newtxtext,newtxmath}
\usepackage{bm}
\usepackage{booktabs}
\usepackage{caption}
\usepackage{xfrac}
\usepackage{placeins}
\usepackage[T1]{fontenc}

\DeclareRobustCommand{\VAN}[3]{#2}
\let\VANthebibliography\thebibliography
\def\thebibliography{\DeclareRobustCommand{\VAN}[3]{##3}\VANthebibliography}

\usepackage{graphicx}	% Including figure files
\usepackage{amsmath}	% Advanced maths commands
\title[Simulation-based inference for cosmic shear with KiDS]{A simulation-based inference pipeline for cosmic shear with the Kilo-Degree Survey}

\author[Lin et al.]{
Kiyam Lin$^{1}$,
Maximilian von Wietersheim-Kramsta$^{1}$,
Benjamin Joachimi$^{1}$,
Stephen Feeney$^{1}$
\\
$^{1}$Department of Physics and Astronomy, University College London, Gower Street, London, WC1E 6BT, UK\\
}

\date{Accepted XXX. Received YYY; in original form ZZZ}

\pubyear{2022}

\begin{document}
\label{firstpage}
\pagerange{\pageref{firstpage}--\pageref{lastpage}}
\maketitle

\begin{abstract}
The standard approach to inference from cosmic large-scale structure data employs summary statistics that are compared to analytic models in a Gaussian likelihood with pre-computed covariance.
To overcome the idealising assumptions about the form of the likelihood and the complexity of the data inherent to the standard approach, we investigate simulation-based inference (SBI), which learns the likelihood as a probability density parameterised by a neural network.
We construct suites of simulated summary statistics, exactly Gaussian-distributed for validation purposes, for the most recent Kilo-Degree Survey (KiDS) weak gravitational lensing analysis and demonstrate that SBI recovers the full 12-dimensional KiDS posterior distribution with just under $10^4$ simulations.
We optimise the simulation strategy by initially covering the parameter space by a hypercube, followed by batches of actively learnt additional points.
The data compression in our SBI implementation is robust to suboptimal choices of fiducial parameter values and of data covariance.
Together with a fast simulator, SBI is therefore a competitive and more versatile alternative to standard inference.
\end{abstract}


%%%%%%%%%%%%%%%%%%%%%%%%%%%%%%%%%%%%%%%%%%%%%%%%%%

%%%%%%%%%%%%%%%%% BODY OF PAPER %%%%%%%%%%%%%%%%%%

\section{Introduction} \label{sec:introduction}

Cosmological weak lensing in the era of modern high-precision cosmology has proven itself to be an excellent probe of key parameters of the standard Lambda Cold Dark Matter Model ($\mathrm{\Lambda CDM}$). Most notably, it is able to constrain a degenerate combination of $\sigma_8$ and $\Omega_\mathrm{m}$, or alternatively $\Omega_\mathrm{m}$ and $S_8$, a combined parameter typically taken as $S_8 = \sigma_8 ( \Omega_\mathrm{m} / 0.3)^{0.5}$. In recent years, two weak lensing surveys, the Kilo-Degree Survey (KiDS, \citealt{asgari2021kids}) and the Dark Energy Survey (DES, \citealt{secco2022dark, amon2022dark}) alongside other photometric galaxy surveys such as the Subaru Hyper Suprim-Cam (HSC, \citealt{sugiyama2022hsc}) have yielded results that are in agreement with each other despite very different methodologies \citep{asgari2021kids, heymans2021kids, secco2022dark, amon2022dark, busch2022kids}. 

Interestingly both KiDS and DES find values of $S_8$ of between 2 to 3$\sigma$ lower than the value inferred by Planck, a space-based experiment observing cosmic microwave background (CMB) anisotropies \citep{ade2016planck, aghanim2020planck}. The consistent results from both KiDS and DES alongside their re-analysis suggests that it is unlikely for the tension to arise simply from some un-modelled systematic error \citep{amon2022consistent}. If this $S_8$ tension cannot be resolved through the discovery of systematic errors, then it motivates the search of new physics.

The analysis of cosmic weak lensing survey data however is fraught with challenges from not only the modelling side but also the limitations of traditional inference methods. To elaborate, the modelling must take into account many factors, such as baryon feedback and the intrinsic alignment of galaxies caused by matter-galaxy interactions \citep{kilbinger2015cosmology, mandelbaum2018weak, amon2022non}. Therefore, we may find a complex statistical problem to solve within the likelihood function, necessary for the task of cosmological parameter inference involving these stochastic forward modelling processes. 

Traditional likelihood analysis requires a likelihood that can be evaluated, but the complete set of these factors makes it impossible to know the exact analytical model of the likelihood written in closed form. For many cases, an accurate likelihood model that takes into account all of the statistical features is essentially too expensive and thus intractable to evaluate \citep{jeffrey2021likelihood}. To this end, it is routine in cosmological surveys to assume a Gaussian likelihood as an approximation to the true likelihood. 

This Gaussian likelihood assumption is employed on summary two-point statistics \citep{asgari2021kids} which are sensitive to the underlying cosmology. However, in the use of two-point statistics such as the correlation function, we may find significant deviations from a Gaussian likelihood when we consider the two-point statistics' sensitivity to low multipoles \citep{schneider2009constrained, sellentin2018skewed}.
This is true even when the underlying lensing fields are Gaussian \citep{sellentin2018insufficiency,  sellentin2018skewed, taylor2019cosmic, upham2021sufficiency}. 
%This means that regardless of how well the likelihood of these two-point correlation functions can be approximated by Gaussian likelihoods, there is still some cosmological information being lost due to this approximation. Furthermore, as surveys enter an era of ever increasing fidelity covering larger patches of the sky, there is a corresponding need to be able to perform better cosmological data analysis that will account for non-Gaussian systematics to avoid biasing the analysis, resulting in incorrect inference.}
Systematic effects could also introduce non-Gaussianity, with their relative importance increasing as surveys become more statistically powerful.
%BJ

As such, to tackle the statistical side of cosmological analysis, there has been a growing number of forward simulation based methods \citep{gupta2018non, fluri2018cosmological, ribli2019weak, taylor2019cosmic, jeffrey2021likelihood, fluri2022full, hahn2022rm}. These seem attractive as they completely circumvent the need of evaluating or working with an explicit or computable form of the likelihood function. This then allows these simulation-based inference methods to fully propagate all of the uncertainties and survey systematics from data to parameters through forward simulation. It should be noted however that simulation-based inference methods are not the only methods that circumvent the use of a Gaussian likelihood, for example, Bayesian Hierarchical Models \citep{porqueres2021lifting, porqueres2021bayesian}.

Most tantalising however is that combining likelihood-free, simulation-based inference (SBI) methods with recent advances in machine learning provides an inference methodology that is not only capable of freeing the analysis pipeline from intractable likelihoods, but also computationally cheaper when paired with a similarly fast simulator. For example, in \citet{alsing2019fast}, only 1,000 simulations were needed to infer a posterior with the same constraints as a long Markov chain Monte Carlo (MCMC) run that required at least 10,000 likelihood evaluations to converge. These techniques have been explored in detail by others with high levels of success \citep{gupta2018non, fluri2018cosmological, ribli2019weak, jeffrey2021likelihood, fluri2022full}.

Notably, \cite{fluri2022full} performed a full $w$CDM analysis of KiDS-1000 weak lensing using deep learning. Despite using the same data set as in this analysis, the \cite{fluri2022full} analysis is concerned with going beyond the standard weak lensing framework by using field-level summary statistics in order to constrain cosmologies beyond the standard $\Lambda$CDM. In that analysis, a graph convolutional neural network is trained on the cosmology dependence of the simulated weak lensing maps with respect to four parameters. Using that as a summary statistic, the parameters are then inferred from the data using approximate Bayesian computation. In this work, the aim is instead to show the feasibility of applying density estimation SBI to a standard $\Lambda$CDM weak lensing analysis with its full complexity in parameter space; such an analysis has not been conducted until now. We achieve this by demonstrating that this methodology allows us to infer all 12 of the standard KiDS-1000 parameters using mock data vectors that are Gaussian drawn to allow for validation.
%BJ

Arguably the idea of SBI began with the seminal work of \citet{rubin1984bayesianly} when describing the process of inferring a posterior in Bayesian analysis from a frequentist perspective. The core idea being that if the parameter values to a typically stochastic forward modelling process generates a data vector identical to the observed data vector, then those parameter values must make up part of the inferred posterior. This in turn means that they have a high probability of being the ``true'' parameter values. In essence, a rejection sampling schema for the posterior.

This idea led to the birth of a class of SBI more commonly known today as approximate bayesian computation (ABC). However, one might immediately notice that for any realistic stochastic process, getting an exact match in the data vector is highly improbable, leading to \citet{pritchard1999population} introducing a notion of closeness to allow for imperfect matching, an idea that is also fraught with problems requiring ever increasingly complex notions of closeness. 

Furthermore, with the low probability of obtaining a data vector that passes any such closeness criteria, one can imagine that the ABC methodology is computationally inefficient, requiring many forward simulations to find just one posterior parameter point. In light of this, many have tried to develop better sampling schemas to increase the efficiency of ABC, but even after many such improvements, \citet{leclercq2018bayesian} estimates that computational efficiency remains low. This ABC method, however, has met high levels of success \citep{pritchard1999population, marin2012approximate, ishida2015cosmoabc, akeret2015approximate, jennings2017astroabc, prangle2017adapting, leclercq2018bayesian, fluri2018cosmological, fluri2022full}.

There is a desire to make use of all of the information available from forward simulation, giving rise to an alternative SBI methodology in the form of density estimation likelihood free inference (DELFI). In this schema, the probability density of the sampling distribution is learnt through the use of neural networks. After evaluating this probability density at a given mock or observed data vector, a likelihood and thus posterior can be recovered. This method has the advantage of not requiring an explicit and often simplified form of a likelihood, and also does not throw away the information from forward simulations that do not produce a data vector that passes any ``closeness'' criteria.

In recent years, DELFI has shown itself to be capable of inferring tight constraints on the final posterior surface with an almost order of magnitude fewer forward simulations than traditional Bayesian methods \citep{papamakarios2016fast, alsing2019fast, jeffrey2021likelihood, hahn2022rm}. In this paper, we apply this method of statistical inference to mock KiDS-1000 cosmic shear data. We validate the method using simplified simulations but with the full set of inferred parameters. We also optimise the method to explore how many simulations are needed in comparison to the number of parameters being varied and inferred. The setup involves a mixture of both data sensitive parameters but also prior driven parameters.

This paper is structured as follows: Section~\ref{sec:sbi} details the specifics of the software used to perform DELFI SBI as well as the compression scheme that is applied onto the cosmological summary statistics. Section~\ref{sec:cosmology} provides an overview of the cosmological setup as well as the test simulations used to generate mock data vectors.  In Sect.~\ref{sec:validation_and_optimisation}, we outline the process of validating our SBI methodology, testing for robustness as well as optimising the method for the number of simulations and the learning process. Importantly, in this section we demonstrate that the method can be easily made robust towards both poor choices of fiducial cosmology as well as being robust to sub-optimal compression.

%----------------------------------------------------------------------------------------------------------------------

\section{Simulation Based Inference} \label{sec:sbi}

\subsection{DELFI} \label{sec:pydelfi}

\begin{figure*}
  \captionsetup{width=0.8\linewidth}
  \centering
  \includegraphics[width=\linewidth]{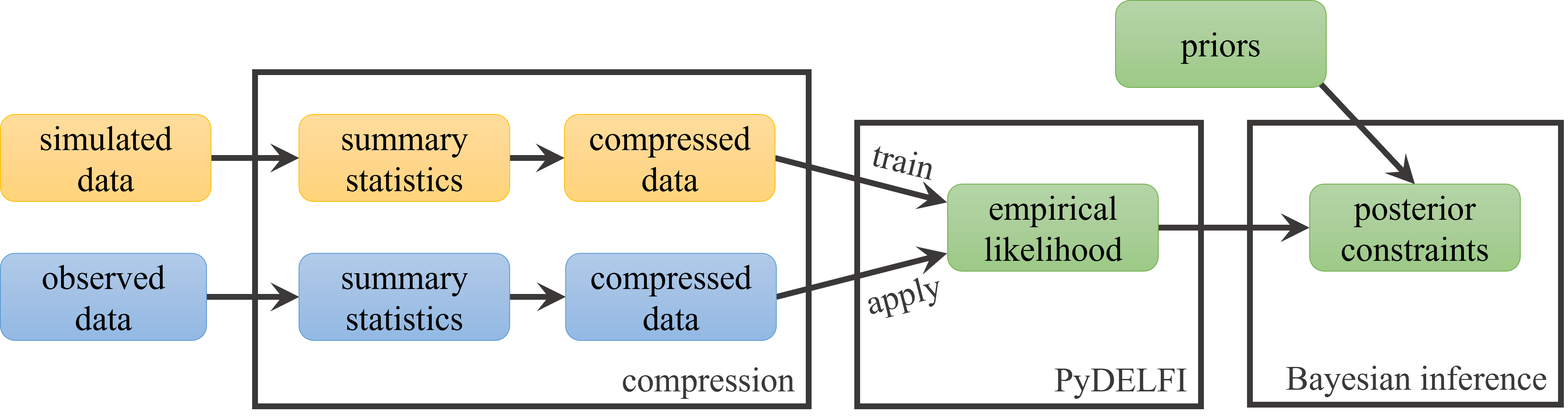}
  \caption{An overview of DELFI SBI using \texttt{PyDELFI}. First, both the observed data and simulated data are compressed into a set of informative summaries. The compressed simulated data is used to train \texttt{PyDELFI}'s neural networks. Applying the compressed observed data yields an empirical likelihood, which after application of the priors produces posterior parameter constraints.}
  \label{fig0:0_sbi_flowchart}
\end{figure*}

The methodology of DELFI SBI is depicted in Fig.~\ref{fig0:0_sbi_flowchart}. First, both the observed data and simulated data are compressed into a set of informative summaries. The compressed simulated data is used to train a neural network, which after applying the compressed observed data yields an empirical likelihood. After multiplying this learned likelihood with the priors, one obtains posterior parameter constrains.

\citet{alsing2019fast} outlined a variety of ways of performing DELFI. The methodology we use involves learning the sampling distribution of data vectors conditional on the input cosmology, $p(d | \theta)$ where $d$ denotes data vector and $\theta$ cosmological parameters. By evaluating this learned sampling distribution at the observed data vector, one obtains the likelihood, which, after multiplication with a prior, yields the posterior through the use of Bayes' theorem \citep{papamakarios2019sequential, lueckmann2019likelihood}.

The biggest advantage of learning the sampling distribution of data as a function of parameters vs. learning the joint distribution is that the networks do not have the prior embedded within them. This means that one can acquire forward simulations in regions of posterior interest without worrying about importance re-weighting issues \citep{papamakarios2019sequential, alsing2019fast}. This methodology also means that different priors can be explored and changed a posteriori without similar re-weighting issues.

The specific implementation of density estimation SBI used is that of the \texttt{PyDELFI} software package\footnote{\url{https://github.com/justinalsing/pydelfi}} \citep{alsing2019fast}. To account for any numerical anomalies or learning problems in the density learning process, a committee of neural density estimators (NDEs) are employed. This means that multiple NDEs are trained independently and later combined, weighted by how well they learned the target density. \texttt{PyDELFI} supports natively two different classes of NDEs, Mixture Density Networks (MDNs) and Masked Autoregressive Flows (MAFs).

We found that MDNs performed poorly for our problem, especially when considering high dimensional parameter spaces, so we will not discuss them further (see \citet{alsing2019fast} for details on the method). We make use of MAFs, which are NDEs constructed out of a chain of Masked Autoregressive Density Estimators (MADEs). As any probability distribution can be written as a chain of one-dimensional conditional probabilities, a MADE effectively learns a target distribution through a series of conditional probability distribution transformations, $p(t_i | \bm{\mathrm{t}}_{1: i-1}, \theta)$, back to the unit normal, where $\bm{\mathrm{t}}$ is the data vector of interest. The means and variances are parameterised by a neural network with weights, $\bm{\mathrm{w}}$ \citep{germain2015made, uria2016neural, alsing2019fast}. A MADE therefore has a functional form of

\begin{equation} \label{eq:neural_prob_chain_rule}
    p(\bm{\mathrm{t}}|\bm{\theta}; \bm{\mathrm{w}}) = \prod^{\mathrm{dim} (\bm{\mathrm{t}})}_{i = 1} p(t_i | \bm{\mathrm{t}}_{1: i-1}, \bm{\theta}; \bm{\mathrm{w}}).
\end{equation}

\noindent where each conditional $p(t_i | \bm{\mathrm{t}}_{1: i-1}, \bm{\theta}; \bm{\mathrm{w}})$ is conditioned on having observed the previous $\bm{\mathrm{t}}_{1: i-1}$ data vector values. \citet{alsing2019fast} write that any single MADE has two key limitations. One limitation is that a single MADE is sensitive to the order of the factorisation whilst the other is that simple conditionals may not be flexible enough to learn complex target distributions. The information pertaining to suitability of conditionals or factorisation order however is not available a priori.

To overcome these limitations, MAFs are employed. A MAF addresses these limitations by creating a stack of individual MADEs, where the output $\bm{\mathrm{u}}$ of each MADE is used as the input distribution for the next MADE in the stack \citep{papamakarios2017masked}. By creating an ensemble of MADEs with random re-ordering of the factorisation order between each MADE, the limitation and sensitivity of factorisation order is overcome. \citet{papamakarios2019sequential} writes that these MAFs are very flexible neural density estimators and well suited to the task of likelihood-free inference. A MAF as a neural density estimator can thus be expressed as

\begin{equation}
    p(\bm{\mathrm{t}}|\bm{\theta}; \bm{\mathrm{w}}) = \mathcal{N}[\bm{\mathrm{u}}(\bm{\mathrm{t}}, \theta; \bm{\mathrm{w}})|\bm{0}, \bm{\mathrm{I}}] \times \prod^{N_{\mathrm{mades}}}_{n=1} \prod^{\mathrm{dim(\bm{t})}}_{i=1} p_i^n(\bm{\mathrm{t}}, \theta; \bm{\mathrm{w}}),
\end{equation}

\noindent where $\bm{\mathrm{u}}$ is the output from the final MADE. $\mathcal{N}[\bm{\mathrm{u}}(\bm{\mathrm{t}}, \theta; \bm{\mathrm{w}})|\bm{0}, \bm{\mathrm{I}}]$ denotes the aforementioned unit normal and $\prod^{\mathrm{dim(\bm{t})}}_{i=1} p_i^n(\bm{\mathrm{t}}, \theta; \bm{\mathrm{w}})$ represents the chain of conditional probabilities factorised via the chain rule. The reason this expression picks up the $\mathcal{N}[\bm{\mathrm{u}}(\bm{\mathrm{t}}, \theta; \bm{\mathrm{w}})|\bm{0}, \bm{\mathrm{I}}]$ term is because we can think of a MADE, or a stack of MADEs in a MAF as learning the transformation of $\bm{\mathrm{t}}$ back to the unit normal as hinted in \citet{alsing2019fast}.

To train these NDEs, \texttt{PyDELFI} minimises the Kullback-Leibler divergence between the parametric density estimator and the target density. However, for the purposes of simulation based inference, the target probability density is not known. As such, a Monte Carlo estimate of the Kullback-Leibler divergence is used instead for the target density. More details can be found in \citet{alsing2019fast}.

However, any application of machine learning trained on a small training set by minimisation of the loss function easily runs into the problem of potentially optimising said density estimators for local minima that do not well represent the larger data set as a whole. As such \texttt{PyDELFI} employs the technique of training an ensemble of NDEs with a range of network architectures. By doing so, one constructs a stack of NDEs, with stacking weights (contribution weighting) of each NDE given by the relative likelihoods for each NDE. A stack of NDEs combined this way is reported to perform better than any single NDE \citep{smyth1998evaluation, smyth1999linearly}. This gives,

\begin{equation} \label{eq:stack_NDEs}
    p(\bm{\mathrm{t}}|\theta; \bm{\mathrm{w}}) = \prod^{N_\mathrm{NDEs}}_{\alpha = 1} \beta_\alpha p_\alpha(\bm{\mathrm{t}}| \theta; \bm{\mathrm{w}}),
\end{equation}

\noindent where $\beta_\alpha$ represents the stacking weight of each NDE with index $\alpha$ and $\sum \beta_\alpha = 1$.

It is possible to run \texttt{PyDELFI} in two different modes which we will refer to as \textit{batch run} mode and \textit{active learning} mode. In the \textit{batch run} mode, simulations are run beforehand before being fed as one batch to \texttt{PyDELFI}. This means that to use the \textit{batch run} mode setting, it is typically prudent to select parameter points at which to run forward simulations by sampling from the prior with an appropriate method, such as by using an equally spaced grid or a latin hypercube. The main drawback of this mode is that with any individual standalone run, it is hard to tell whether a sufficient number of forward simulations have been run without some ground truth to which the results can be compared. This is because the absolute value of the loss is dependent on the number of parameters being inferred, and there is not one target loss to aim for across all models and runs. 

In the active learning mode, \texttt{PyDELFI} proposes new parameter values at which to run simulations after obtaining data vectors from a small initial set of simulations. This means that for the initial set of simulations one can choose to either randomly sample from the prior or make use of a latin hypercube that maximally covers the prior volume efficiently. We choose the latter. After training on this initial set of simulations, \texttt{PyDELFI} then acquires further sets of parameter-data pairs by sampling from a weighted mixture of the intermediate posterior and the prior, however, other parameter acquisition schemes may be used.

The performance of neural networks is typically also sensitive to their initialisation. For our work, given that we know that the likelihood will be approximately Gaussian around its peak from the published results of the KiDS-1000 team \citep{asgari2021kids}, we can make use of a Fisher matrix, $\bm{\mathrm{F}}$, multiplied by a factor of safety to initialise our ensemble of NDEs. We express the Fisher matrix as

\begin{equation}
    \bm{\mathrm{F}} = \nabla \bm{\mu}^\mathrm{T} \bm{\mathrm{C}}^{-1} \nabla^\mathrm{T} \bm{\mathrm{\mu}},
\end{equation} 

\noindent where $\nabla \bm{\mu}$ is the derivative of the data vector at the fiducial cosmology and $\bm{\mathrm{C}}$ the data covariance assuming a Gaussian likelihood (See \citet{alsing2018generalized} for more details). We have made use of the fact that the covariance is cosmology independent here. 

The factor of safety is introduced to ensure that the NDEs are not initialised with too restrictive a volume. In practice, this means we initialise our NDEs with a Gaussian target distribution with their means equal to the chosen fiducial cosmology parameter values and a covariance equal to the inverse Fisher matrix multiplied by a constant acting as a factor of safety. To elaborate, the NDEs are initialised before training to $p(\bm{\mathrm{t}}| \bm{\theta}) = \mathcal{N}(\bm{\mathrm{t}} | \bm{\theta}, k \bm{\mathrm{F}}^{-1})$, where $k$ denotes the aforementioned factor of safety constant.

\subsection{Parameter sampling} \label{sec:parameter_sampling}

The process of SBI means that ideally the forward simulations that generate the data vectors are drawn from a set of parameters that maximally cover the prior volume. As such a latin hypercube covering the prior volume would be ideal as it is a method of maximally covering the volume of parameters of interest with minimal computational waste by not sampling any particular parameter values twice \citep{stein1987large, park1994optimal, loh1996latin}. As the KiDS prior contains a mix of top hat and Gaussian components, the hypercube generating algorithms provided by the \texttt{PyDOE}\footnote{\url{https://github.com/tisimst/pyDOE}} package could be employed alongside \texttt{SciPy}\footnote{\url{https://github.com/scipy/scipy}} with only minor modifications. 

Step one of the algorithm is to divide the prior volume into a lattice of equally spaced hyper-cuboids. Step two is to then randomly choose hyper-cuboids such that along each dimension no parameter interval is sampled twice. From here, either a parameter point can be picked randomly within the chosen intervals, or the centre of each interval can be chosen. Running the inference pipeline on either choice appears to make little difference, so for simplicity's sake the middle of each interval is chosen as the sampled value for our work. This random choosing of hyper-cuboids can be run multiple times. A measure of minimum Euclidean distance between any two points is used as a measure of how spread out the points are within the hypercube, and any random sampling that produces a larger minimum Euclidean distance is deemed to be a better sample. This process can be repeated an arbitrary number of times with the only caveat being an increase in computational cost.

The process ends here for parameters that have a flat prior, but for parameters that have a Gaussian prior there is an additional step. For the parameters that have a Gaussian prior, \texttt{SciPy} is used to map the flat spread of parameter points onto a Gaussian target distribution through its inverse cumulative distribution function.

\subsection{Score compression} \label{sec:score_compression}
The data vector we have chosen for this work is that of weak lensing two-point correlation functions, which in the KiDS-1000 setup has length 270 (see Sec.~\ref{sec:cosmology}). Using data vectors of such length directly within \texttt{PyDELFI} would be prohibitively expensive and difficult to fit due to its high dimensionality. Hence, massive data compression is necessary to compress the data vectors into highly informative summary statistics with minimal loss in information.

For this massive data compression task, there are a few methods to choose from. One option would be to use a neural network to try and maximise the information content automatically. For example, this can be done through the use of the software package \texttt{IMNN} \citep{charnock2018imnn}, which uses a neural network to construct summaries that maximise the Fisher information. Another data compression method is MOPED \citep{heavens2000massive}, a linear compression method built upon the more classic method of Karhunen-Lo\'eve eigenvalue decomposition \citep{tegmark1997karhunen}. However, since we know from previous analyses that the parameter likelihood from cosmic shear two-point statistics will be approximately Gaussian close to the peak of the likelihood \citep{schneider2009constrained, sellentin2018skewed, sellentin2018insufficiency, taylor2019cosmic, upham2021sufficiency}, we make use of linear score compression as outlined in \citet{alsing2018generalized}.

Given a log-likelihood, $\mathcal{L}$, its Taylor expansion around a set of fiducial parameters, $\bm{\theta}_*$ with respect to $\delta \bm{\theta}$ can be written as,

\begin{equation} \label{eq:log_likelihood}
    \mathcal{L} = \mathcal{L}_* + \delta \bm{\theta}^\mathrm{T} \nabla \mathcal{L}_* - \frac{1}{2} \delta \bm{\theta}^\mathrm{T} \mathbf{J}_* \delta \bm{\theta},
\end{equation}

\noindent where a $_*$ denotes evaluation at the fiducial parameter values, $\mathbf{J}_*$ is the observed information matrix, $\mathbf{J}_* \equiv - \nabla \nabla^T \mathcal{L}_*$. To linear order in the parameters, the data vector only couples to the parameters via the $\nabla \mathcal{L}_*$ term, commonly referred to as the score, with $\mathbf{s} \equiv \nabla \mathcal{L}$. By construction this score function is a vector of length $n$, where $n$ is the number of parameters. This is as derivatives of the log-likelihood are taken with respect to $\bm{\theta}$. As such, this provides a natural method for compressing any data vector of length $N$ to a massively smaller vector of length $n$, sensitive to changes in the parameters to the first order. \citet{alsing2018generalized} remark that this linear compression method generalises the linear Karhunen-Lo\'eve compression and MOPED schemas considered in \citet{heavens2000massive} and \citet{tegmark1997karhunen}.

However, as already alluded to above, this method of compression does require some form of an approximate likelihood with respect to the parameters of interest: the more accurate the likelihood, the more optimal the compression. It should be emphasised that an inaccurate likelihood approximation would only result in lossy compression, and would not bias the inference, simply producing wider posterior contours. For our work, we choose a Gaussian form for the likelihood for which this method of compression saturates the Cram\'er-Rao bound \citep{alsing2018generalized}. 

% This yields a functional form of the score of

% \begin{equation} \label{eq:all_term_score}
%     \bm{s} = \nabla\bm{\mu}^\mathrm{T}\bm{C}^{-1}(\bm{d} - \bm{\mu}) + \frac{1}{2} (\bm{d} - \bm{\mu})^\mathrm{T} \bm{C}^{-1}  \nabla\bm{C}\bm{C}^{-1} (\bm{d} - \bm{\mu}) - \frac{1}{2} tr(\bm{C}^{-1}\nabla \bm{C}),
% \end{equation}

% \noindent where $\bm{d}$ represents the data vector, $\bm{\mu}$ the fiducial mean of data vector and $\bm{C}$ the covariance matrix. Assuming that $\bm{\mu} = \bm{\mu}(\theta)$ and $\bm{C} = \bm{C}(\theta)$, i.e. both the means and the covariance are dependent on parameters. We have also made use of the following relationships: $\nabla \bm{C}^{-1} = \bm{C}^{-1}  \nabla\bm{C}\bm{C}^{-1}$ and $\nabla \ln|\bm{C}| = tr(\bm{C}^{-1}\nabla \bm{C})$.

% In Eq.~(\ref{eq:all_term_score}), the last term does not couple with the data vector $\bm{d}$ in any way, and so can be discarded for the purposes of score compression. 

Moreover, we assume that the covariance is parameter-independent, which allows us to drop any partial derivatives with respect to the covariance. The contribution to the Fisher matrix by the parameter dependence of the covariance is suppressed for larger survey area (REF Tegmark, Taylor, Heavens). We note that the SBI analysis does fully account for the parameter dependence of statistical errors, as opposed to the standard Gaussian likelihood analysis.
%This choice is safe for the future analysis of the real KiDS-1000 data as the covariance has low sensitivity to the parameters \citep{joachimi2021kids}.}
%BJ
This leaves us with a greatly simplified linear score compression function, defined as,

\begin{equation} \label{eq:score_compression}
    \bm{t} = \nabla\bm{\mu}^\mathrm{T}\bm{C}^{-1}(\bm{d} - \bm{\mu}),
\end{equation}

\noindent where $\bm{t}$ now denotes the compressed data that will be used as the information-sufficient summary statistics to be fed into \texttt{PyDELFI}. 

However, with the expression of score compression as given in Eq.~(\ref{eq:score_compression}), the numerical values of the compressed summary statistics are not directly informative. Through Eq.~(\ref{eq:log_likelihood}), \cite{alsing2018generalized} show that by maximising the Taylor-expanded log-likelihood, the score compression numbers can be mapped onto a quasi maximum-likelihood estimator through

\begin{equation} \label{eq:mle_fisher}
    \hat{\bm{\theta}} = \bm{\theta}_* + \bm{F}_*^{-1} \nabla \mathcal{L}_* = \bm{\theta}_* + \bm{F}_*^{-1} \bm{t}_*,
\end{equation}

\noindent where $\bm{F}_*$ denotes the Fisher matrix evaluated at the fiducial cosmology and $\bm{\theta}_*$, the fiducial cosmology parameter values. We use these quasi maximum-likelihood estimators as the summary statistics within \texttt{PyDELFI} to be able to make use of the Fisher initialisation schema mentioned in Sect.~\ref{sec:pydelfi}.

Furthermore, it is in this compression step that we may choose to marginalise over certain parameters such as any nuisance parameters. For example, if we know that our data is only sensitive to a subset of the parameters varied through forward simulation, those parameters can be marginalised out of the analysis through the score. \citet{alsing2019nuisance} proposed a method of doing this whilst maximising the information within the data vector's sensitivity to those parameters through use of the Fisher matrix, a method they call ``nuisance hardened score compression''. 

For our cosmological setup, the parameters that we might choose to marginalise out contain little information, meaning that the Fisher matrix yielded little to no information pertaining to the marginalised parameters. This meant that for our work, we could marginalise our compressed summaries by simply truncating the score, removing the terms that corresponded to the parameters we wished to marginalise over. To mirror the analysis done by the KiDS-1000 team however, no marginalisation of the score is done throughout this work. It should be noted that if the covariance depended on parameters, then such a clean separation between statistic and parameter is impossible, and instead more mixing would be observed between compressed statistics and parameter values. This would also result in a more complex marginalisation process beyond simply truncating the score.

%---------------------------------------------------------------------------

\section{Simulations of cosmic shear data from KiDS-1000} \label{sec:cosmology}

For the generation of mock data from initial parameters, forward simulation is done using the \texttt{KCAP}\footnote{\url{https://github.com/KiDS-WL/kcap}} and  \texttt{CosmoSIS}\footnote{\url{https://bitbucket.org/joezuntz/cosmosis/wiki/Home}} packages \citep{zuntz2015cosmosis}. For this analysis, the output of the simulations are the shear two point correlation functions (2PCFs), $\xi_{\pm}(\theta)$. To calculate the 2PCFs, firstly the cosmological pipeline assumes a spatially flat $\mathrm{\Lambda CDM}$ model. The linear matter power spectrum is calculated using \texttt{CAMB} \citep{lewis2000efficient, howlett2012cmb} with its non-linear evolution calculated using \texttt{HMCode} \citep{mead2015accurate}. \texttt{HMCode} makes use of a halo model with baryonic feedback. The amplitude of the halo mass-concentration, $a_\mathrm{bary}$, is allowed to vary freely, whilst the halo model bloating parameter $\eta_0$ is fixed in relation to $a_\mathrm{bary}$ (see \citet{joachimi2021kids} for more details).

The effects of the intrinsic alignment of galaxies is factored in through the non-linear alignment model of \citet{bridle2007dark}. Following the pipeline as set out by KiDS \citep{asgari2021kids}, the Limber approximation is used to project the matter power spectrum along the line of sight to obtain $C_{\epsilon \epsilon}(\ell)$, which are the observed cosmic shear angular power spectrum that are dependent on multipole $\ell$. The total shear angular power spectrum is the sum of contributions from gravitational lensing (G) and intrinsic alignments (I), giving,

\begin{equation}
    C_{\epsilon \epsilon}(\ell) =  C_{\mathrm{GG}}(\ell) +  C_{\mathrm{GI}}(\ell) +  C_{\mathrm{II}}(\ell).
\end{equation}

\noindent The $C_{\epsilon \epsilon}(\ell)$ are subsequently transformed into 2PCFs, $\xi{\pm}$,

\begin{equation}
    \xi{\pm} (\theta) = \int^{\infty}_{0} \frac{\mathrm{d}\ell \, \ell}{2 \pi} \mathrm{J}_{0/4}( \ell \theta) C_{\epsilon \epsilon}(\ell) ,
\end{equation}

\noindent with $\mathrm{J}_{0/4}$ denoting Bessel functions of the first kind and $\theta$ the angular separation on the sky. Following KiDS-1000, we assume zero contribution from the B-modes to the 2PCFs.

The source galaxies are split up into five redshift bins with bin boundaries [0.1, 0.3, 0.5, 0.7, 0.9]. All of the cross-correlation and auto-correlation pairs between respective redshift bins were taken into account resulting in 15 redshift bin pairs. Following KiDS-1000, scale cuts are performed on the 2PCFs, only keeping angular separations of between 0.5 and 300 arcminutes with a total of 9 angular bins. This resulted in a data vector of length 270.

As the goal is to test the performance of the \texttt{PyDELFI} SBI pipeline, we generate our own mock data vectors by sampling the 2PCFs using the KiDS analytic covariance as derived in \citet{joachimi2021kids}. We do this to have full control over the results and to perform valid testing of the SBI methodology by way of comparison to that of traditional likelihood analysis. Furthermore, for the purposes of testing we generated a mock data vector that was used as the observed data vector for both traditional likelihood analysis and SBI. 

In the future, we plan to apply this pipeline to a novel suite of physically informed forward-simulations of weak lensing observables (von Wietersheim-Kramsta, Lin, et al., in prep.). For this full SBI analysis of KiDS-1000 data, the simulations will include all relevant systematics and physical effects which might induce non-Gaussianity into the likelihood. 

Following the methodology of KiDS, we vary five cosmological parameters and two astrophysical nuisance parameters with flat priors identical to the ones used by KiDS (see section~3.3 of \citet{joachimi2021kids}). The cosmological parameters varied include $\sigma_8$, the present day root-mean-square matter fluctuation averaged over a sphere of radius $8h^{-1} \mathrm{Mpc}$; the density parameter for cold dark matter, $\omega_\mathrm{c} = \Omega_\mathrm{c} h_0^2$ and baryonic matter, $\omega_\mathrm{b} = \Omega_\mathrm{b} h_0^2$ multiplied by $h_0$, the dimensionless Hubble constant. The spectral index of the primordial power spectrum, $n_s$, is likewise varied. 

The two astrophysical nuisance parameters are $A_{\mathrm{IA}}$, the intrinsic alignment amplitude of galaxies and $a_{\mathrm{bary}}$, the baryonic feedback amplitude. Furthermore, we define matter density as $\Omega_\mathrm{m} = \Omega_\mathrm{c} + \Omega_\mathrm{b} + \Omega_\mathrm{\nu}$, where $\Omega_\mathrm{\nu}$ is the neutrino density. Finally, the shifts in the means of the redshift distribution bins follow a Gaussian prior with covariance that can be found in the latest KiDS data release repository\footnote{\url{https://github.com/KiDS-WL/Cat\_to\_Obs\_K1000\_P1/}}. For our analysis, the mean shift in the redshift distribution is set to zero. See Table~\ref{tab:theta_vals} for a summary of the parameters varied and their prior ranges. 

It should be noted that the 2PCFs are only strongly sensitive to the parameters $\sigma_8$, $\omega_\mathrm{c}$ and $A_{\mathrm{IA}}$. This means that we expect the prior to dominate the posterior for all of the other parameters that are varied.

\begin{table} 
    \centering
    \begin{tabular}{|l|c|l|l|}
        \toprule 
        Parameter           & Prior Range               & Mock Data & Fiducial  \\
        \midrule
        $\sigma_8$          &\ [0.6, 1.0] \             &\ 0.8 \    &\ 0.811    \\
        $\omega_\mathrm{b}$ &\ [0.019, 0.026] \         &\ 0.0230 \ &\ 0.0224   \\
        $\omega_\mathrm{c}$ &\ [0.07, 0.18] \           &\ 0.120 \  &\ 0.120    \\
        $n_\mathrm{s}$      &\ [0.8, 1.15] \            &\ 0.960 \  &\ 0.965    \\
        $h_0$               &\ [0.6, 0.9] \             &\ 0.674 \  &\ 0.674    \\
        $a_\mathrm{bary}$   &\ [2.0, 4.0] \             &\ 3.10 \   &\ 3.13     \\
        $A_\mathrm{IA}$     &\ [-6.0, 6.0] \            &\ 0.960 \  &\ 0.974    \\ 
        $\delta_z [5]$      &\ $\mathcal{N}(\mu, C)$    &\ 0.0      &\ 0.0      \\   
        \bottomrule
    \end{tabular}
    \caption{The prior ranges for the parameters to be inferred alongside the parameter values used to generate the mock data vector and the fiducial cosmology for compression. The prior ranges for $\sigma_8$,$\omega_\mathrm{b}$, $\omega_\mathrm{c}$, $n_\mathrm{s}$, $h_0$, $a_\mathrm{bary}$ and $A_\mathrm{IA}$ are all top hats, whilst the prior for $\delta_z$ follows a correlated Gaussian distribution characterised by a covariance matrix $C$ with mean of $\mu$. The $\delta_z$ parameters encapsulate freedom in the mean of the redshift distribution bins whilst the other parameters are: $\sigma_8$, the root-mean-square matter fluctuation; $\omega_\mathrm{b}$, baryonic matter density; $\omega_\mathrm{c}$, cold dark matter density; $n_\mathrm{s}$, scalar spectral index; $h_0$, Hubble constant; $a_\mathrm{bary}$, baryonic feedback parameter; $A_{\rm{IA}}$, galaxy intrinsic alignment amplitude. We set the equation of state parameter as $w=-1$, pick a flat curvature, $\omega_\mathrm{k} = 0$, and fix a neutrino mass sum of $\Sigma m_\mathrm{\nu} = 0.06 \mathrm{eV}/c^2$.}
    \label{tab:theta_vals}
\end{table}

\section{Validation \& Optimisation} \label{sec:validation_and_optimisation}

\subsection{SBI methodology validation} \label{sec:validation}

To validate the methodology, using the setup described in Sect.~\ref{sec:cosmology}, both a mock observed data vector and fiducial cosmology data vector were generated. We choose Planck 2018 cosmology values \citep{aghanim2020planck} as our fiducial cosmology but pick a slightly different set of cosmology values to generate the mock observed data vector. We test robustness and sensitivity to the choice of fiducial cosmology later in Sect.~\ref{sec:sensitivity_to_fiducial}.

The generated data vectors are compressed following the schema outlined in Sect.~\ref{sec:score_compression}. The compressed summary statistics are then fed into a \texttt{PyDELFI} pipeline that initialises the NDEs with the inverse Fisher matrix as mentioned in Sect.~\ref{sec:pydelfi} before training the ensemble of NDEs on forward simulated data-parameter pairs.

Importantly, as our data is only sensitive to a subset of the parameters, with the other parameters being highly prior-driven, we add this prior information to the inverse Fisher matrix used for NDE initialisation via the method set out in \cite{coe2009fisher}. This will have no effect on the final inferred parameter posterior, but it helps regularise the NDE initialisation making it perform more consistently.

In the end, the chosen ensemble of NDEs used included six MAFs comprised of three to eight MADEs, respectively. This choice was made after trying a variety of different combinations of NDEs, with this combination providing good performance without being too restrictive as would be the case if MAFs with fewer components were chosen.

\begin{figure*}
  \captionsetup{width=0.8\textwidth}
  \centering
  \includegraphics[width=\textwidth]{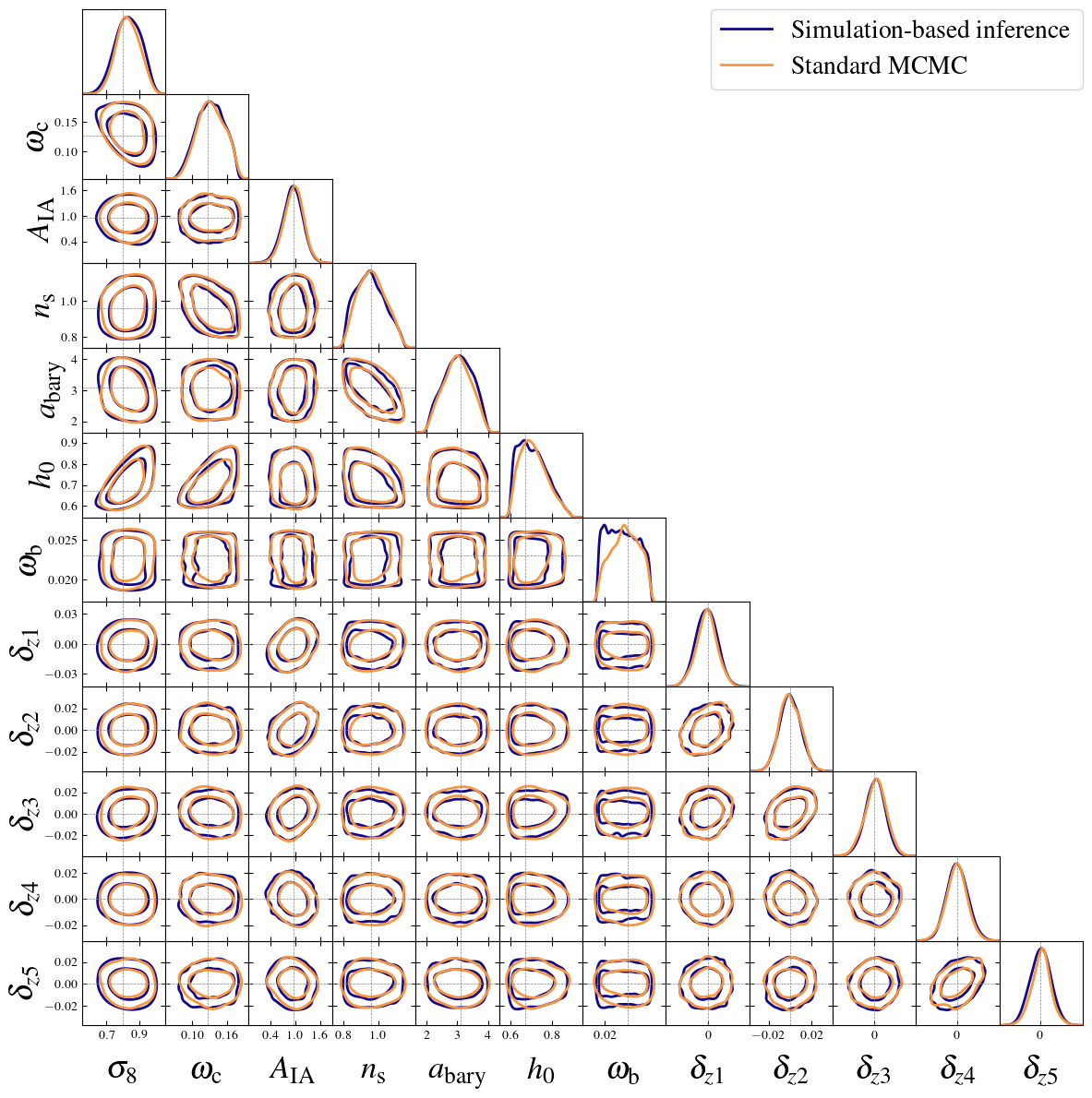}
  \caption{Posterior distributions of the full parameter set of KiDS-1000 obtained through a standard MCMC analysis (orange) vs. the SBI pipeline (blue). Choices for the fiducial cosmology and the cosmology parameter values for the mock observed data are outlined in Table~\ref{tab:theta_vals}. The dashed grey lines depict the cosmology used to generate the mock observed data. The SBI contours were obtained with \texttt{PyDELFI} in its active learning mode that made use of 11,000 forward simulations. In comparison, the MCMC analysis made use of 48,000 model evaluations.}
  \label{fig1:1_pydelfi_vs_emcee_all_parameters}
\end{figure*}

The results of the SBI pipeline were compared each time to a standard likelihood inference pipeline using \texttt{emcee}\footnote{\url{https://github.com/dfm/emcee}} \citep{foreman2013emcee} that made use of 48,000 model evaluations, with convergence tested using the integrated correlation time as recommended by \cite{foreman2013emcee}. As mentioned previously, the results of the standard likelihood inference pipeline were treated as the ground truth for testing purposes. Figure~\ref{fig1:1_pydelfi_vs_emcee_all_parameters} shows the comparison between the traditional likelihood analysis vs. the output of the SBI pipeline after running forward simulations with all 12 parameters varied. It is clear that the SBI pipeline is able to reproduce the ground truth posterior with all 12 cosmological parameters varied. 

This particular posterior was obtained after 11,000 forward simulations with \texttt{PyDELFI} set to run in its active learning mode. The one-dimensional marginal posteriors differ slightly for $h_0$ and $\omega_\mathrm{b}$, which is due to the posterior being prior driven with a low sensitivity to the data. This means that we expect the posterior to be flat for these parameters. 

In particular for $\omega_\mathrm{b}$, we can see that the posterior from SBI reflects the expected flat distribution better than the standard MCMC analysis that assumed a Gaussian likelihood. This shows us that the SBI methodology accurately reflects any deficiency in information within the data vector concerning parameter constraints.

Classifier two-sample tests (C2ST) are often used to determine how well a posterior has been learned, whereby a classifier is trained to see if it can distinguish between samples from the ground-truth distribution and samples from the learned distribution \citep{friedman2003multivariate, lopez2016revisiting}. A value of 0.5 in the test would indicate the classifier cannot distinguish between the two distributions whilst a v alue of 1.0 would indicate the classifier can perfectly distinguish between the two distributions. We found that our methodology when tested with C2ST was competitive with what \cite{Miller2021truncated} found the performance of \texttt{PyDELFI} to be, giving a value of 0.65 with $\mathcal{O}(10^4)$ simulations when only considering three-dimensional marginalised posteriors in $\sigma_8, \Omega_\mathrm{m}, A_\mathrm{IA}$ and a value of 0.6 when only a two-dimensional posterior is considered.

\subsection{SBI sensitivity to choice of fiducial cosmology} \label{sec:sensitivity_to_fiducial}

Our SBI methodology requires a choice of fiducial cosmology both to initialise the NDEs as well as to do score compression. There is, however, no way to know what a good choice of fiducial cosmology is a priori, and importantly we would not want the choice of fiducial cosmology to bias the results. Instead, a poor choice of fiducial cosmology should only result in sub-optimal compression. Thus, it was important to test this methodology for robustness against choice of fiducial cosmology.

To test for this, we generated a set of 100 mock observed data vectors using varying cosmologies that spanned the prior in $\sigma_8$ and $\Omega_\mathrm{m}$. For this set of mock observed data vectors, the other parameter values were kept fixed to the values depicted in Table~\ref{tab:theta_vals}. For the sake of simplicity, instead of running the SBI pipeline in its active learning mode, a batch of 24,000 forward simulations were pre-run with their data vectors compressed using the schema outlined in Sect.~\ref{sec:score_compression} and all 12 parameters varied drawn from a latin hypercube. Separate runs of the SBI pipeline were then performed using each of the individual mock observed data vectors.

\begin{figure}
  \captionsetup{width=\linewidth}
  \centering
  \includegraphics[width=\linewidth]{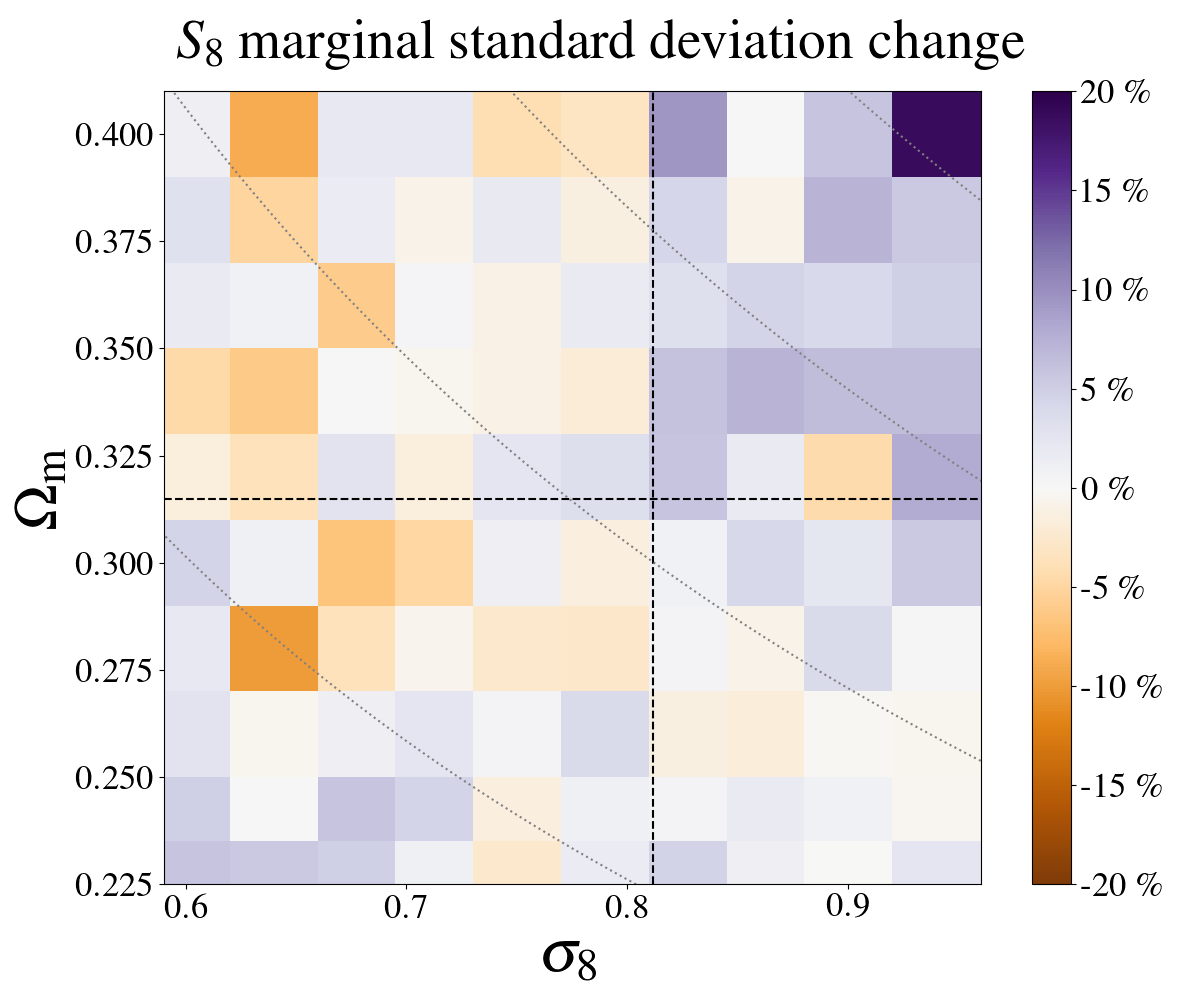}
  \caption{The relative size of the $S_8$ marginal posterior standard deviation compared to a fiducial analysis where the mock data vector was the same as the mock data vector. The $\Omega_\mathrm{m}$ and $\sigma_8$ axes depict the cosmology used to generate the mock data vector, whilst the colour maps the percentage difference in the standard deviation of the $S_8$ marginal posterior. The black dashed lines depict the fiducial cosmology values, whilst the dotted grey lines that span the figure diagonally depict lines of constant $S_8$. The standard deviation of the $S_8$ marginal posterior is up to 20\% different to the case where the data cosmology aligned with the fiducial cosmology.}
  \label{fig3:3_s_8_deltas}
\end{figure}

We first found that the inferred posterior is always consistent with the cosmology used to generate the mock observed data vectors, even when the fiducial cosmology lay outside of the posterior. We then wished to see how the standard deviation in $S_8$ is affected by the choice in fiducial cosmology. Figure~\ref{fig3:3_s_8_deltas} depicts the percentage change in standard deviation in the $S_8$ marginal posterior. We find that when the true $S_8$ and corresponding $\sigma_8$ value is larger than the fiducial $S_8$ and $\sigma_8$ value, the posterior in $S_8$ is artificially widened, whilst for the rest of parameter space in $\Omega_\mathrm{m}$ and $\sigma_8$ there is no clear trend. For the majority of cases, the change in standard deviation is under 5\%; a small percentage for an inherently noisy process due to both cosmic variance changing with $S_8$ and the stochastic nature of NDE training. This indicates to us that after running an initial analysis with a fiducial cosmology that will yield parameter constraints consistent with the data cosmology, it would be prudent to re-compress the data vector with the newly inferred data cosmology. % change made here

In practice this first involves performing inference with a fiducial cosmology. The maximum a posteriori (MAP) of this inference can be found with an optimiser such as Nelder-Mead \citep{nelder1965simplex}, and we use the MAP parameter values to re-compress our data. We then re-perform \texttt{PyDELFI} inference with this once-iterated MAP cosmology to yield more accurate constraints. 

\begin{figure}
  \captionsetup{width=0.8\linewidth}
  \centering
  \includegraphics[width=\linewidth]{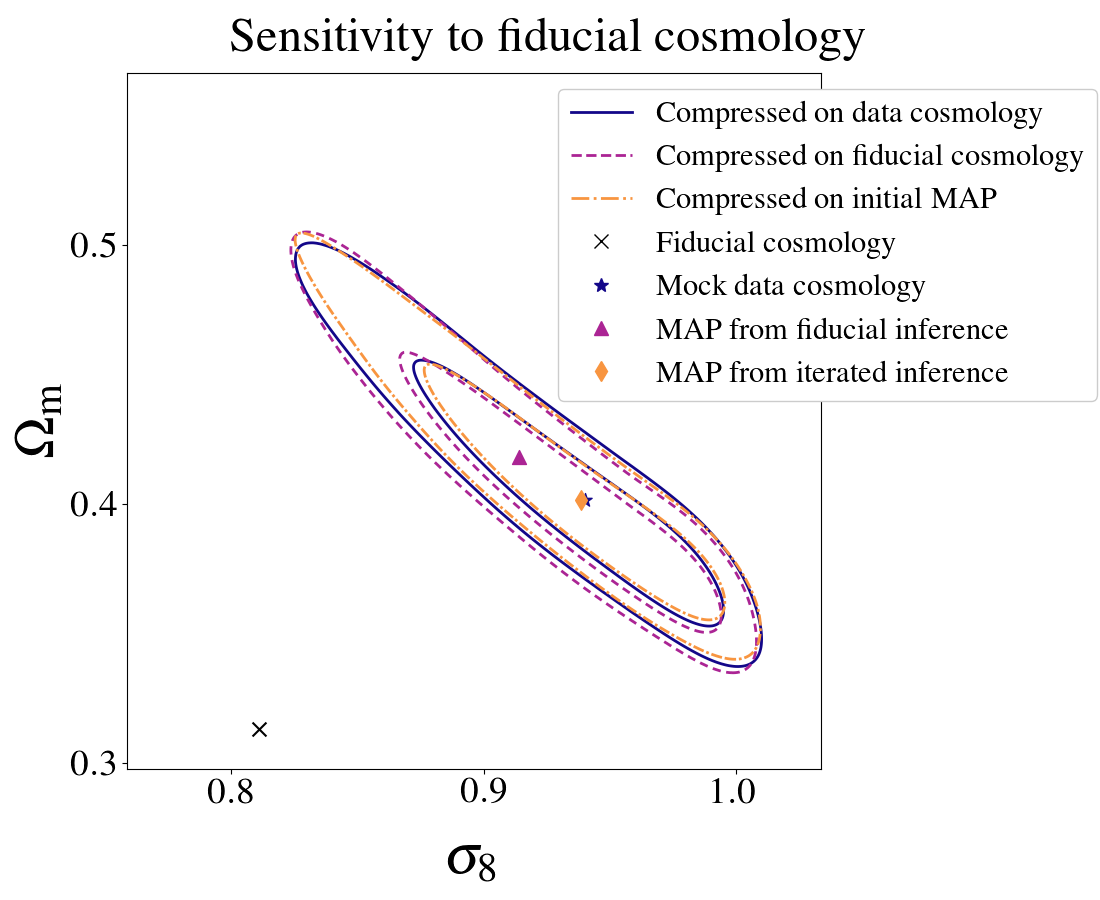}
  \caption{SBI analysis where the mock observed data was generated from a cosmology (blue star) that deviated far from the choice of fiducial cosmology (black cross). The maximum a posteriori (MAP) was found from this initial fiducial inference (fuschia triangle) with corresponding posterior contours (fuchsia dashed line), and used to re-compress the data vectors which resulted in new posterior contours (yellow dash dotted line). The MAP from this once iterated inference is depicted by the yellow diamond. Inference performed with summaries that were compressed with the cosmology used to generate the mock observed data is shown by the solid blue line.}
  \label{fig2:2_data_sensitivity_in_banana}
\end{figure}

Figure~\ref{fig2:2_data_sensitivity_in_banana} depicts this process using the cosmology depicted by the top right hand corner in Fig.~\ref{fig3:3_s_8_deltas}, where the difference in $S_8$ standard deviation between inference performed with compression on the fiducial cosmology and inference performed with compression on the mock data cosmology was 19\%. This was the worst-case scenario that we encountered in our testing. After compressing the data on the fiducial cosmology, we infer the MAP and re-perform the compression to obtain new posterior contours with a $S_8$ standard deviation that is now only 5\% different to that of inference performed with compression on the mock data cosmology. This process can be iterated several times if required. Furthermore, all of the MAP values have very similar $S_8$, but the MAP from the once-iterated inference is also very close to the true $\sigma_8$ and $\Omega_\mathrm{m}$ values.

It should be noted that re-performing the \texttt{PyDELFI} inference with a new cosmology is computationally inexpensive as we can make use of all of the previously run simulations to perform the fiducial inference. There is only a small amount of computational cost associated with calculating data vector derivatives with respect to cosmology at the first iteration MAP, and also to re-calculate the score compression followed by training a new set of neural density estimators. As such, we would always recommend to perform at least one iteration and see if the MAP or standard deviation varies significantly, and to continue iterating until neither the MAP nor the standard deviation vary by much depending on the amount of noise present. This result shows that the SBI methodology can easily be made robust towards the choice of fiducial cosmology after just one inference iteration.

\subsection{SBI sensitivity to quality of compression}

As discussed in Sect.~\ref{sec:score_compression}, a crucial step in the SBI pipeline is the compression. Whilst for our testing we were able to make use of optimal compression by using the same covariance to both draw data vector values and in the compression, this will not generally be the case. This is as an analytical covariance may not always be available and a covariance matrix would also be unable to completely capture the non-Gaussian features of forward simulated data. Therefore, there is a need to test the methodology in lieu of lossy compression, which we do by tampering with the covariance used in compression, artificially worsening it.

We tamper with the covariance in two ways. In the first method, as our analytic sample data covariance matrix follows a Wishart distribution \citep{wishart1928generalised, taylor2013putting}, we draw random samples of the covariance from a Wishart distribution constructed from our data covariance whilst varying the degrees of freedom. By reducing the degrees of freedom, we increase the amount of noise in the covariance like if we were estimating the covariance numerically from data samples, with lower degrees of freedom corresponding to fewer data samples. In the second method of tampering with the covariance, we suppress the off-diagonal elements of the covariance by a factor of $10^{-l}$, where $l$ denotes the diagonal distance from the diagonal element. This has the effect of destroying all of the cross correlation information within the compression. This approach of tampering the covariance was chosen to ensure that the covariance structure would be significantly compromised but retaining its positive definiteness; it does not mimic any physical effect in the covariance modelling.

\begin{figure*}
  \captionsetup{width=0.8\linewidth}
  \centering
  \includegraphics[width=\linewidth]{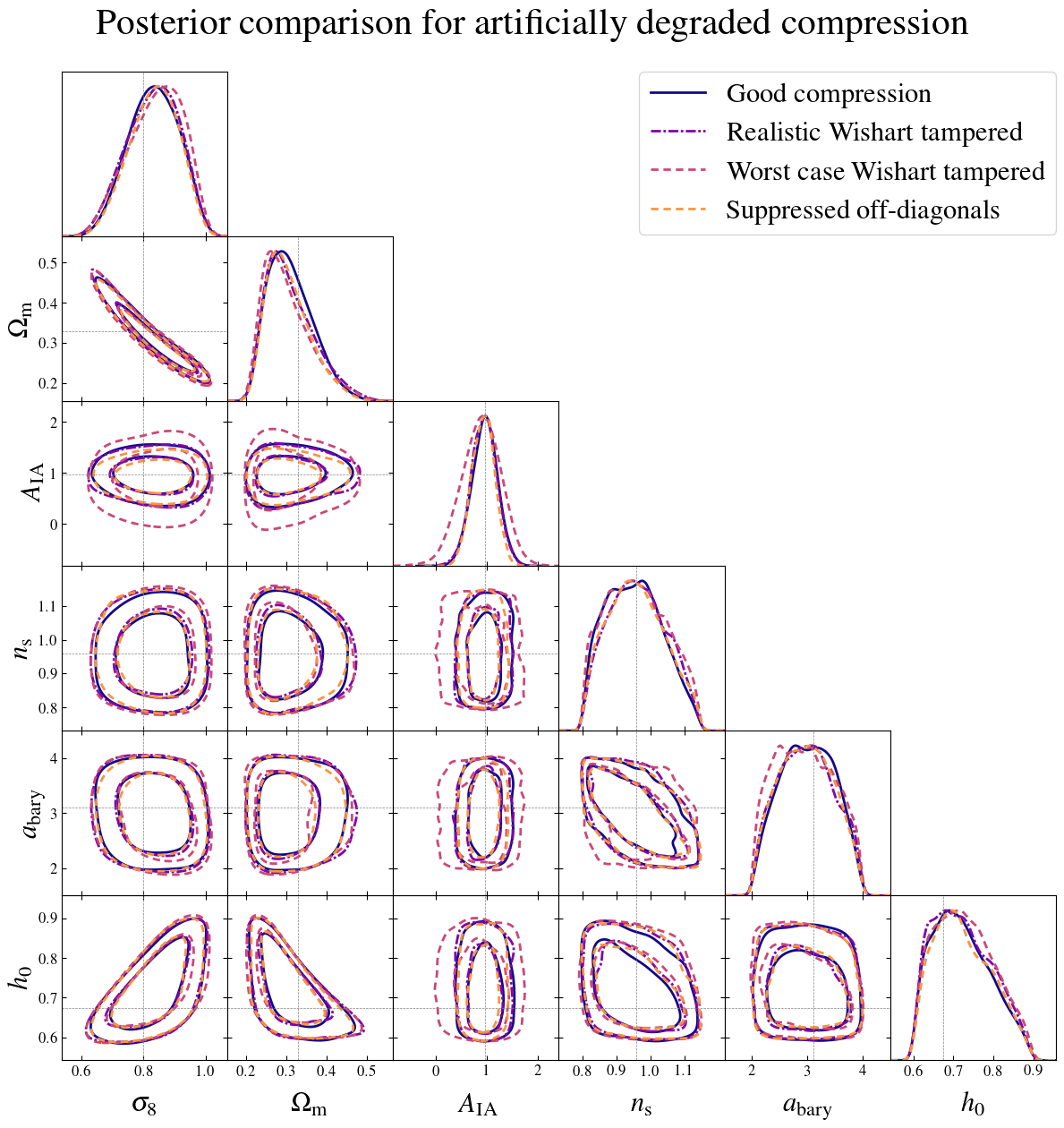}
  \caption{Posterior contours obtained through the SBI pipeline using three artificially worsened compression covariances plotted on top of the posterior obtained from using a good compression covariance (solid blue). In the realistic Wishart tampered case (purple dash dotted), a covariance for use in compression was resampled from a Wishart distribution with the degrees of freedom set to 1000, a reasonable number of forward simulations one might perform to estimate a data covariance. In the worst case Wishart tampered case (pink dashed), the Wishart degrees of freedom was set to 312, just above the limit of the degrees of freedom for our data covariance that is 270 to keep the matrix invertible \citep{taylor2013putting}. In the suppressed off-diagonals case (orange dashed), the covariance had its off-diagonal elements suppressed by a factor of $10^{-l}$, where $l$ denotes the diagonal distance from the diagonal element. This strongly suppresses all of the cross-correlation information within the compression. The dashed grey lines depict the cosmology used to generate the mock observed data. It should be noted that in all but the worst Wishart tampered case where only the $A_\mathrm{IA}$ contours widen, the SBI method is able to obtain a posterior almost identical to the good compression case.}
  \label{fig4:4_pydelfi_tampered_compression_posteriors}
\end{figure*}

Figure~\ref{fig4:4_pydelfi_tampered_compression_posteriors} depicts the posterior contours obtained through the SBI pipeline using these artificially worsened compression methods. We can see from this figure that in all but the worst Wishart tampered case where the degrees of freedom was set to 312, just above the degrees of freedom limit of 270 to keep the covariance matrix invertible, the posteriors obtained through SBI do not differ greatly from the case with good compression. In the realistic Wishart tampered case, the degrees of freedom was set to 1000, a reasonable number of forward simulations one might perform to estimate a data covariance. However, even in the worst case scenario, we can see that the effect is almost entirely posterior widening and mostly only on the $A_\mathrm{IA}$ parameter.

It is clear that this SBI pipeline is robust towards lossy compression and can infer good posteriors under such circumstances. This sensitivity test also indicates that making use of a numerically estimated, and thus noisy covariance would suffice for purposes of re-analysing KiDS-1000.

\subsection{Optimisation} \label{sec:optimisation}

We wish to know how many potentially expensive forward simulations are required to successfully make use of this SBI methodology. We find that the active learning approach requires fewer simulations to produce a well learned and stable posterior. However, there are downsides: the networks go through more rounds of training overall, as they are retrained each time a new set of simulations is acquired. In terms of training speed however, retraining \texttt{PyDELFI} NDEs is faster than even the simplified simulations that we are making use of here. This indicates that they will be much faster than the more realistic simulations we will make use of in the re-analysis of KiDS-1000 data. To elaborate, the simulations we are running as outlined in Sect.~\ref{sec:cosmology} take place on the order of a minute per realisation whilst training an entire \texttt{PyDELFI} model on a modern CPU with no GPU acceleration takes around 10 to 15 minutes. This shows us that the speed of analysis is dominated by the time it takes to run the forward simulations.

As we have mentioned previously in Sect.~\ref{sec:parameter_sampling}, we draw initial parameter points at which to run the forward simulations using a latin hypercube. We found that the key benefit of using a latin hypercube is that the tails of the parameter space are well explored. However, just relying on a latin hypercube led to the peaks of the posterior lacking the number of simulations required to converge. As such, the active learning mode of \texttt{PyDELFI} draws further parameter points at which to run forward simulations from a weighted mix of the intermediate posterior and prior. We wished to see the impact on the training by drawing parameter points in this manner, hence we compared the results of running the active learning approach against a single large latin hypercube with the same number of total simulations.

\begin{figure}
  \captionsetup{width=0.8\linewidth}
  \centering
  \includegraphics[width=\linewidth]{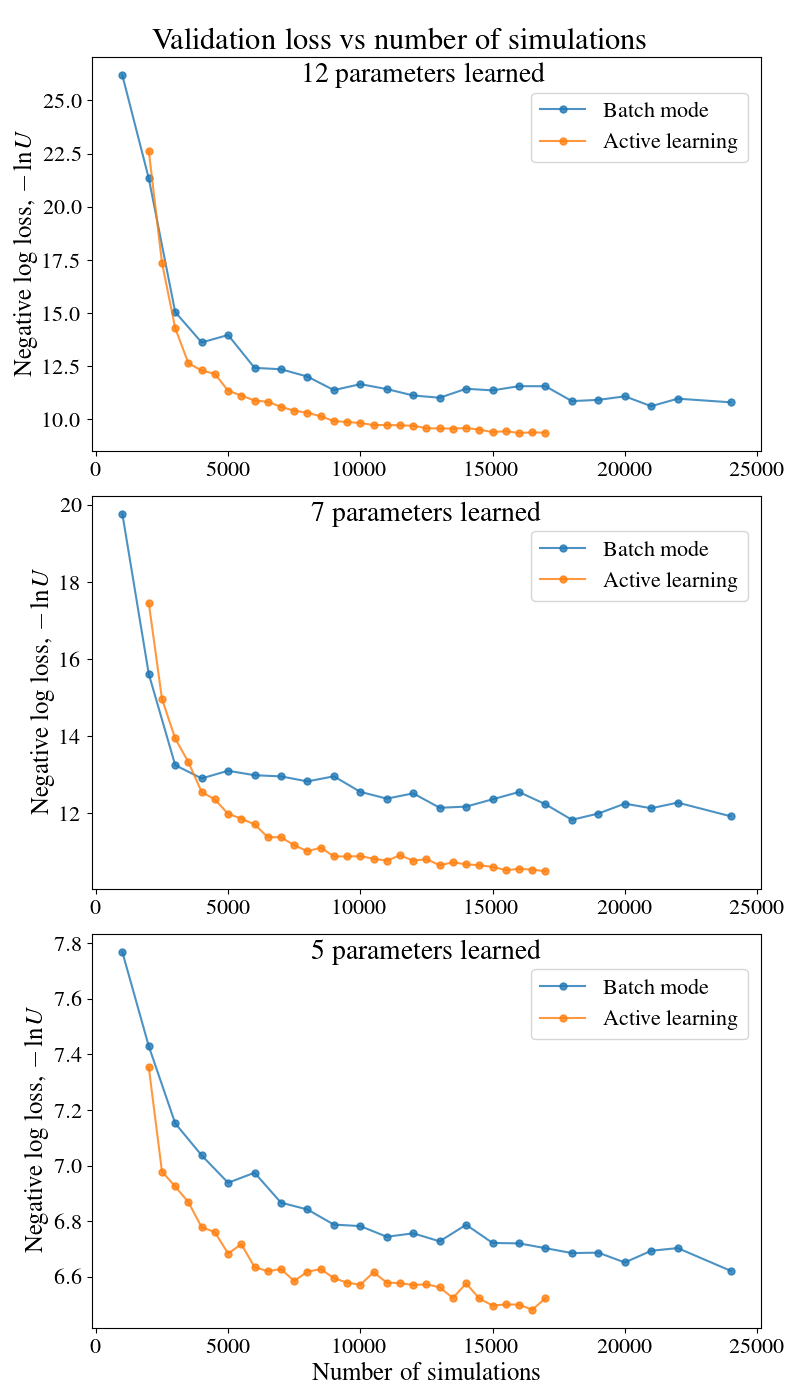}
  \caption{The plot depicts validation loss against the number of simulations for the two modes that \texttt{PyDELFI} can be run in. The orange points are the validation loss from \textit{active learning} whilst the blue points depict the \textit{batch run} mode, both modes are outlined in Sect.~\ref{sec:pydelfi}. For all simulations, all twelve parameters were varied. In the top panel \texttt{PyDELFI} also inferred all twelve parameters, whilst in the middle panel \texttt{PyDELFI} only inferred seven parameters, which were \{$\sigma_8$, $\omega_\mathrm{c}$, $A_\mathrm{IA}$, $a_\mathrm{bary}$, $n_\mathrm{s}$, $h_0$, $\omega_\mathrm{b}$\}. The bottom panel depicts a scenario where \texttt{PyDELFI} inferred five parameters, which were \{$\sigma_8$, $\omega_\mathrm{c}$, $A_\mathrm{IA}$, $a_\mathrm{bary}$, $n_\mathrm{s}$\}. For more than $\mathcal{O}(10^4)$ simulations, the gain in information becomes minimal, with the exact number of simulations dependent on the number of parameters inferred.}
  \label{fig5:5_stacked_validation_loss_active_learning_vs_hypercube}
\end{figure}

There are several metrics that we can use to determine the number of forward simulations we need to obtain good posterior contours. One method is to look at the spread in the learned $S_8$ posterior across each of the individual NDEs in the ensemble that was employed by \texttt{PyDELFI}. Another method is to check if the marginal $S_8$ posterior's standard deviation has converged. We found that both of these metrics were not strongly conclusive, and regardless of the number of simulations, neither metric appreciably deteriorated. As an alternative, we turned to the validation loss of the neural networks to see when the loss stopped decreasing when increasing the number of simulations. Figure~\ref{fig5:5_stacked_validation_loss_active_learning_vs_hypercube} depicts such loss curves, comparing both the validation loss from the active learning mode as well as a latin hypercube with the same number of samples. We find that the active learning approach always outperforms the latin hypercube batch run mode. 

To test for the number of simulations required however, we can look at where the steepness of the log loss curves starts to plateau. When the log loss plateaus, we can deduce that the networks are no longer able to glean much information from adding on further simulations. We can see that for the 12-parameter case, it is around 10,000 simulations that the loss curve starts to plateau. For the 7-parameter case this takes place around 8,000 simulations whilst for 5 parameters closer to 7,000. This shows us that the number of simulations required scales with the number of parameters being inferred. However, as the scaling of simulations required vs. parameters inferred is not polynomial, it is relatively cheap to infer more parameters.

\begin{figure*}
  \captionsetup{width=0.8\linewidth}
  \centering
  \includegraphics[width=\linewidth]{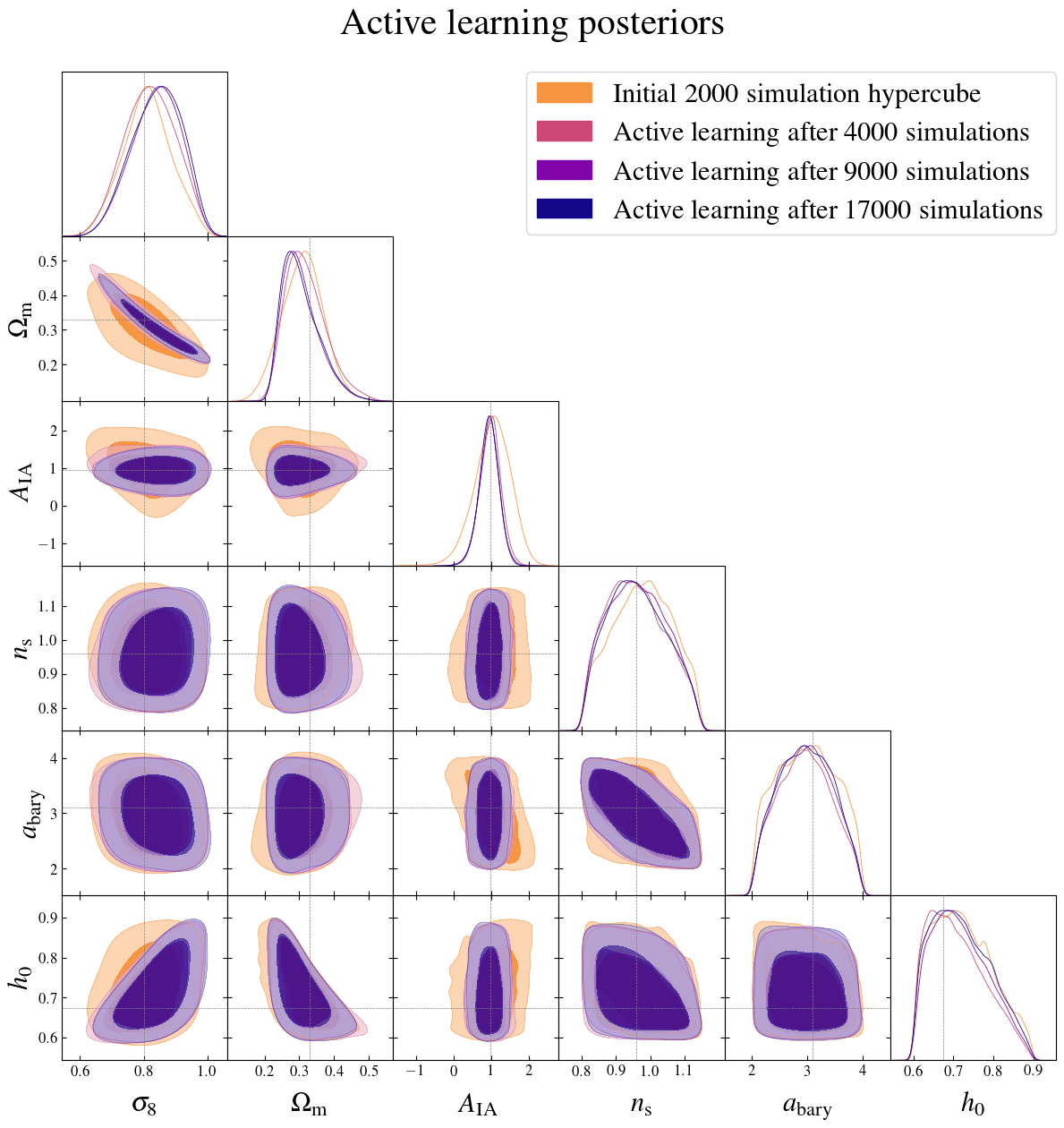}
  \caption{Intermediate posteriors from the active learning mode in \texttt{PyDELFI} for 2000 hypercube samples (orange), 4000 samples (fuchsia), 9000 samples (purple) and 17000 samples (blue). The dashed grey lines depict the cosmology used to generate the mock observed data. The posterior is poorly approximated after the initial 2,000 simulations, yet improves rapidly with just a doubling of simulation number, and after 9000 simulations is almost identical to the posterior obtained after 17000 simulations.}
  \label{fig6:6_snl_intermediate_delfi_posteriors}
\end{figure*}

Furthermore, it is of note that after the initial hypercube, the active learning method rapidly improves its quality of training. To graphically depict this, Fig.~\ref{fig6:6_snl_intermediate_delfi_posteriors} plots the intermediate posteriors overlaid on top of each other. The $\delta_z$ parameters have been marginalised out just to make the plot more visually clear, but they were still learned in this particular run. We can see from this plot that the posterior quickly converges with small increases in simulation number, starting from a poor posterior constraints with the 2,000 parameter points hypercube. Yet, the 9,000 simulation intermediate posterior is almost identical to the one obtained after 17,000 forward simulations. This serves to further reinforce what we see in the loss curves depicted in Fig.~\ref{fig5:5_stacked_validation_loss_active_learning_vs_hypercube}.

Within the setup of \texttt{PyDELFI} itself however, we also tried tuning other hyper-parameters, such as the number of epochs, i.e. the number of training rounds the NDEs undergo; the learning rate, a parameter that determines the amount the NDEs changes after each training epoch; the early stopping threshold, a parameter that determines when to stop training to guard from over-fitting. We found the settings that \texttt{PyDELFI} uses by default to be good, with tweaks providing little to no improvement. The only hyper-parameter we changed was the number of epochs. For high dimensionality inference, the number of epochs needed to be set high enough to give the NDEs enough training rounds to learn the features of the data fully.

Throughout this optimisation testing, the full set of KiDS-1000 parameters was varied; however, we also tried marginalising out the other parameters in the compression step to see if that made the learning task easier. This is prudent to try as it both lowers the dimensionality of the problem and also guided by our knowledge of the cosmological setup, we know that many of the parameters are almost purely prior-driven. For our setup however, given that there was little sensitivity to the data in the parameters that we wished to marginalise, performing nuisance hardened compression as outlined in \cite{alsing2019nuisance} made no difference.

If the \texttt{MOPED} compression schema was used in its original Gram-Schmidt orthogonalisation form \citep{heavens2000massive}, then the sampling distribution of summary statistics would have a covariance structure of that of a unitary matrix. This in principle should be more straightforward for the NDEs to learn, but would require modifications to the current initialisation schema that makes use of a Fisher matrix requiring the summary statistics to be cast into quasi maximum likelihood estimators.

\begin{figure*}
  \captionsetup{width=0.8\linewidth}
  \centering
  \includegraphics[width=\linewidth]{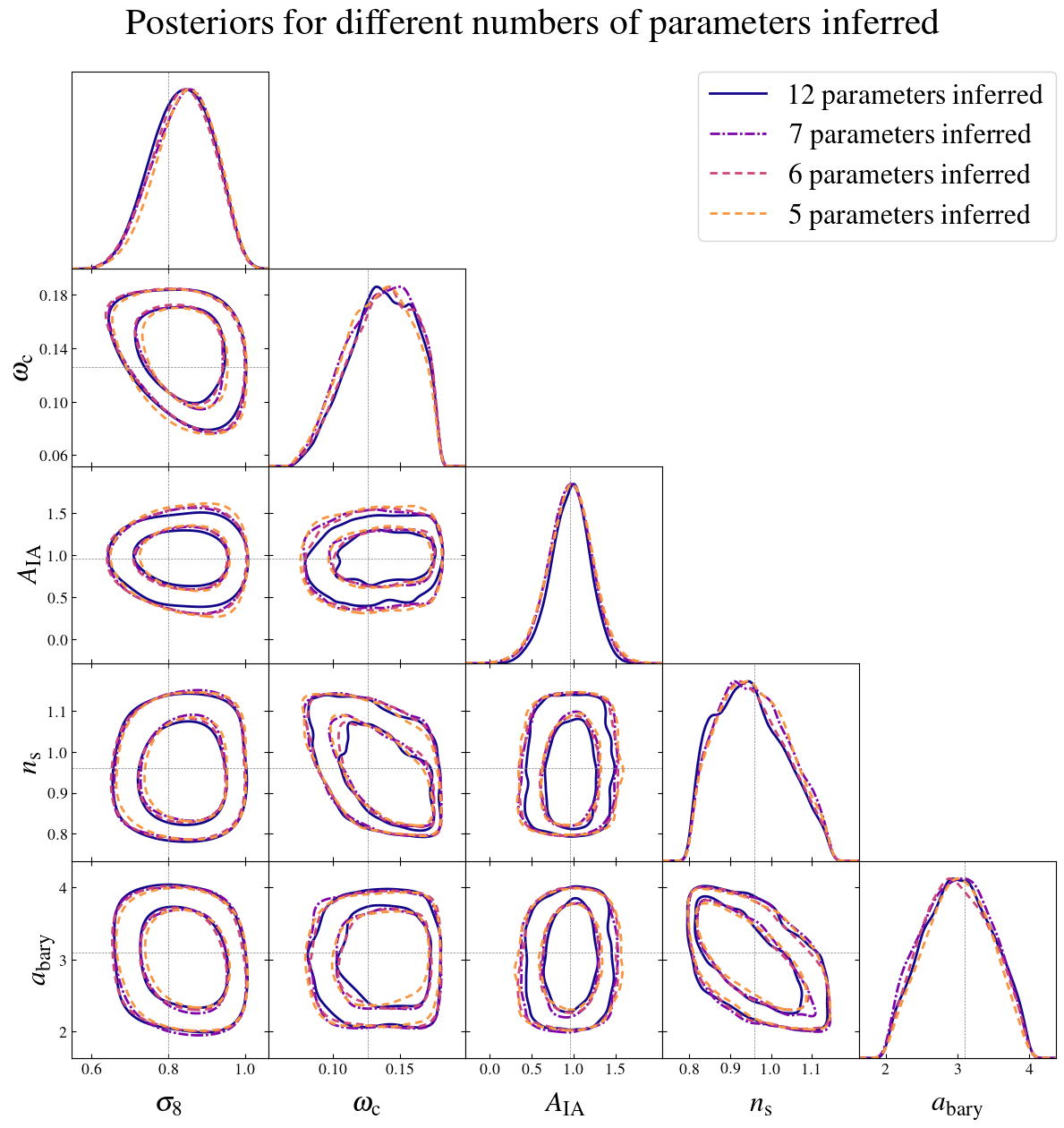}
  \caption{Marginal posteriors from the active learning mode in \texttt{PyDELFI} for varying numbers of parameters inferred. All twelve parameters were varied in the forward simulations, and also \texttt{PyDELFI} inferred all twelve parameters in the solid blue contour. The dashed grey lines depict the cosmology used to generate the mock observed data. For the 7-parameter inference (dash dotted purple), the $\delta_z [5]$ parameters were not inferred, the 6-parameter inference (fuchsia dashed) also dropped $\omega_\mathrm{b}$ and the 5-parameter inference further dropped $h_0$.}
  \label{fig8:8_varied_number_of_parameters_inferred}
\end{figure*}

We also wished to test if reducing the number of parameters inferred would artificially narrow or widen the constraints on the parameters being inferred. We found that there was no such effect on the final parameter posteriors, meaning it is safe to reduce the number of parameters inferred but keep them varied in the forward simulations. The marginal posteriors for this testing are depicted in Fig.~\ref{fig8:8_varied_number_of_parameters_inferred}.

\section{Conclusions}

% Brief summary
As we enter an era of high-precision cosmology, it will become increasingly difficult to reliably extract information from complex data via analytic models and likelihoods. Therefore, we explored simulation-based inference (SBI) as a methodology that would enable forward-modelling and avoidance of a Gaussian likelihood assumption as is common in most cosmological analyses. We tested SBI on the full 12-dimensional parameter space of the most recent Kilo-Degree Survey cosmological analysis of tomographic weak gravitational lensing data (KiDS-1000), assuming a Gaussian data vector to validate the SBI methodology, employing density estimation likelihood-free inference (DELFI) using the \texttt{PyDELFI} software package.

% key results
We demonstrated that our SBI method accurately recovers the full cosmological posterior of the KiDS-1000 analysis when applied to a mock data vector drawn from a Gaussian likelihood. This was achieved with under $10^4$ forward simulations. Moreover, we showed that the necessary maximal data compression step in our method is robust to employing an inaccurate data covariance, and readily made robust towards the choice of fiducial parameter values. This suggests that our approach will still perform well when using approximate analytic covariances or noisy numerical covariance estimates.

Furthermore, we found the most computationally efficient mode in which to run \texttt{PyDELFI} to be an initial latin hypercube of parameter values followed by additional batches determined by \textit{active learning}.
Marginalising parameters that are varied in the forward simulations in the score compression also does not bias the parameter constraints on the remaining parameters. This allows the number of forward simulations required to be further reduced if certain parameters are not of interest for the posterior.

% Interpretation
The tests for robustness we have performed for our SBI method show that it is  competitive for performing accurate parameter inference all whilst dropping the Gaussian likelihood assumption or being restricted to analytic models of the data. If paired with fast yet comprehensive simulations, SBI inference will also not dramatically increase computational requirements.
In forthcoming work we will back-end our SBI pipeline with realistic forward simulations of weak lensing and apply this inference pipeline to KiDS-1000 (von Wietersheim-Kramsta, Lin, et al., in prep.).
While we have restricted our analysis to two-point statistics to be able to validate against the standard inference approach, SBI is readily applicable to any combination of summary statistics that are accurately represented in the simulations, which makes it a powerful tool to extract maximal amounts of information from next-generation cosmological surveys that contain more non-Gaussian information.

\section*{Acknowledgements}

This work was partially enabled by funding from the UCL Cosmoparticle Initiative. MWK thanks the Science and Technology Facilities Council (STFC) for support in the form of a PhD Studentship. BJ acknowledges support by STFC Consolidated Grant ST/V000780/1. For the purpose of open access, the authors have applied a creative commons attribution (CC BY) licence to any author-accepted manuscript version arising.

%%%%%%%%%%%%%%%%%%%%%%%%%%%%%%%%%%%%%%%%%%%%%%%%%%
\section*{Data Availability}

The code will be made publicly available upon acceptance. The data underlying this article will be shared on reasonable request to the corresponding author.

\newpage



\bibliographystyle{mnras}
\bibliography{bibliography} % if your bibtex file is called example.bibs

\begin{thebibliography}{}
\makeatletter
\relax
\def\mn@urlcharsother{\let\do\@makeother \do\$\do\&\do\#\do\^\do\_\do\%\do\~}
\def\mn@doi{\begingroup\mn@urlcharsother \@ifnextchar [ {\mn@doi@}
  {\mn@doi@[]}}
\def\mn@doi@[#1]#2{\def\@tempa{#1}\ifx\@tempa\@empty \href
  {http://dx.doi.org/#2} {doi:#2}\else \href {http://dx.doi.org/#2} {#1}\fi
  \endgroup}
\def\mn@eprint#1#2{\mn@eprint@#1:#2::\@nil}
\def\mn@eprint@arXiv#1{\href {http://arxiv.org/abs/#1} {{\tt arXiv:#1}}}
\def\mn@eprint@dblp#1{\href {http://dblp.uni-trier.de/rec/bibtex/#1.xml}
  {dblp:#1}}
\def\mn@eprint@#1:#2:#3:#4\@nil{\def\@tempa {#1}\def\@tempb {#2}\def\@tempc
  {#3}\ifx \@tempc \@empty \let \@tempc \@tempb \let \@tempb \@tempa \fi \ifx
  \@tempb \@empty \def\@tempb {arXiv}\fi \@ifundefined
  {mn@eprint@\@tempb}{\@tempb:\@tempc}{\expandafter \expandafter \csname
  mn@eprint@\@tempb\endcsname \expandafter{\@tempc}}}

\bibitem[\protect\citeauthoryear{Ade et~al.,}{Ade et~al.}{2016}]{ade2016planck}
Ade P.~A.,  et~al., 2016, Astronomy \& Astrophysics, 594, A13

\bibitem[\protect\citeauthoryear{Aghanim et~al.,}{Aghanim
  et~al.}{2020}]{aghanim2020planck}
Aghanim N.,  et~al., 2020, Astronomy \& Astrophysics, 641, A6

\bibitem[\protect\citeauthoryear{Akeret, Refregier, Amara, Seehars  \&
  Hasner}{Akeret et~al.}{2015}]{akeret2015approximate}
Akeret J.,  Refregier A.,  Amara A.,  Seehars S.,   Hasner C.,  2015, Journal
  of Cosmology and Astroparticle Physics, 2015, 043

\bibitem[\protect\citeauthoryear{Alsing \& Wandelt}{Alsing \&
  Wandelt}{2018}]{alsing2018generalized}
Alsing J.,  Wandelt B.,  2018, Monthly Notices of the Royal Astronomical
  Society: Letters, 476, L60

\bibitem[\protect\citeauthoryear{Alsing \& Wandelt}{Alsing \&
  Wandelt}{2019}]{alsing2019nuisance}
Alsing J.,  Wandelt B.,  2019, Monthly Notices of the Royal Astronomical
  Society, 488, 5093

\bibitem[\protect\citeauthoryear{Alsing, Charnock, Feeney  \& Wandelt}{Alsing
  et~al.}{2019}]{alsing2019fast}
Alsing J.,  Charnock T.,  Feeney S.,   Wandelt B.,  2019, Monthly Notices of
  the Royal Astronomical Society, 488, 4440

\bibitem[\protect\citeauthoryear{Amon \& Efstathiou}{Amon \&
  Efstathiou}{2022}]{amon2022non}
Amon A.,  Efstathiou G.,  2022, Monthly Notices of the Royal Astronomical
  Society, 516, 5355

\bibitem[\protect\citeauthoryear{Amon et~al.,}{Amon
  et~al.}{2022a}]{amon2022consistent}
Amon A.,  et~al., 2022a, arXiv preprint arXiv:2202.07440

\bibitem[\protect\citeauthoryear{Amon et~al.,}{Amon
  et~al.}{2022b}]{amon2022dark}
Amon A.,  et~al., 2022b, Physical Review D, 105, 023514

\bibitem[\protect\citeauthoryear{Asgari et~al.,}{Asgari
  et~al.}{2021}]{asgari2021kids}
Asgari M.,  et~al., 2021, Astronomy \& Astrophysics, 645, A104

\bibitem[\protect\citeauthoryear{Bridle \& King}{Bridle \&
  King}{2007}]{bridle2007dark}
Bridle S.,  King L.,  2007, New Journal of Physics, 9, 444

\bibitem[\protect\citeauthoryear{Busch et~al.,}{Busch
  et~al.}{2022}]{busch2022kids}
Busch J.,  et~al., 2022, Astronomy and Astrophysics, 664, 1

\bibitem[\protect\citeauthoryear{Charnock, Lavaux  \& Wandelt}{Charnock
  et~al.}{2018}]{charnock2018imnn}
Charnock T.,  Lavaux G.,   Wandelt B.~D.,  2018, Astrophysics Source Code
  Library, pp ascl--1804

\bibitem[\protect\citeauthoryear{Coe}{Coe}{2009}]{coe2009fisher}
Coe D.,  2009, arXiv preprint arXiv:0906.4123

\bibitem[\protect\citeauthoryear{Fluri, Kacprzak, Refregier, Amara, Lucchi  \&
  Hofmann}{Fluri et~al.}{2018}]{fluri2018cosmological}
Fluri J.,  Kacprzak T.,  Refregier A.,  Amara A.,  Lucchi A.,   Hofmann T.,
  2018, Physical Review D, 98, 123518

\bibitem[\protect\citeauthoryear{Fluri, Kacprzak, Lucchi, Schneider, Refregier
  \& Hofmann}{Fluri et~al.}{2022}]{fluri2022full}
Fluri J.,  Kacprzak T.,  Lucchi A.,  Schneider A.,  Refregier A.,   Hofmann T.,
   2022, Physical Review D, 105, 083518

\bibitem[\protect\citeauthoryear{Foreman-Mackey, Hogg, Lang  \&
  Goodman}{Foreman-Mackey et~al.}{2013}]{foreman2013emcee}
Foreman-Mackey D.,  Hogg D.~W.,  Lang D.,   Goodman J.,  2013, Publications of
  the Astronomical Society of the Pacific, 125, 306

\bibitem[\protect\citeauthoryear{Friedman}{Friedman}{2003}]{friedman2003multivariate}
Friedman J.~H.,  2003, Statistical Problems in Particle Physics, Astrophysics,
  and Cosmology, 1, 311

\bibitem[\protect\citeauthoryear{Germain, Gregor, Murray  \&
  Larochelle}{Germain et~al.}{2015}]{germain2015made}
Germain M.,  Gregor K.,  Murray I.,   Larochelle H.,  2015, in International
  Conference on Machine Learning. pp 881--889

\bibitem[\protect\citeauthoryear{Gupta, Matilla, Hsu  \& Haiman}{Gupta
  et~al.}{2018}]{gupta2018non}
Gupta A.,  Matilla J. M.~Z.,  Hsu D.,   Haiman Z.,  2018, Physical Review D,
  97, 103515

\bibitem[\protect\citeauthoryear{Hahn et~al.,}{Hahn et~al.}{2022}]{hahn2022rm}
Hahn C.,  et~al., 2022, arXiv preprint arXiv:2211.00660

\bibitem[\protect\citeauthoryear{Heavens, Jimenez  \& Lahav}{Heavens
  et~al.}{2000}]{heavens2000massive}
Heavens A.~F.,  Jimenez R.,   Lahav O.,  2000, Monthly Notices of the Royal
  Astronomical Society, 317, 965

\bibitem[\protect\citeauthoryear{Heymans et~al.,}{Heymans
  et~al.}{2021}]{heymans2021kids}
Heymans C.,  et~al., 2021, Astronomy \& Astrophysics, 646, A140

\bibitem[\protect\citeauthoryear{Howlett, Lewis, Hall  \& Challinor}{Howlett
  et~al.}{2012}]{howlett2012cmb}
Howlett C.,  Lewis A.,  Hall A.,   Challinor A.,  2012, Journal of Cosmology
  and Astroparticle Physics, 2012, 027

\bibitem[\protect\citeauthoryear{Ishida et~al.,}{Ishida
  et~al.}{2015}]{ishida2015cosmoabc}
Ishida E.~E.,  et~al., 2015, Astronomy and Computing, 13, 1

\bibitem[\protect\citeauthoryear{Jeffrey, Alsing  \& Lanusse}{Jeffrey
  et~al.}{2021}]{jeffrey2021likelihood}
Jeffrey N.,  Alsing J.,   Lanusse F.,  2021, Monthly Notices of the Royal
  Astronomical Society, 501, 954

\bibitem[\protect\citeauthoryear{Jennings \& Madigan}{Jennings \&
  Madigan}{2017}]{jennings2017astroabc}
Jennings E.,  Madigan M.,  2017, Astronomy and computing, 19, 16

\bibitem[\protect\citeauthoryear{Joachimi et~al.,}{Joachimi
  et~al.}{2021}]{joachimi2021kids}
Joachimi B.,  et~al., 2021, Astronomy \& Astrophysics, 646, A129

\bibitem[\protect\citeauthoryear{Kilbinger}{Kilbinger}{2015}]{kilbinger2015cosmology}
Kilbinger M.,  2015, Reports on Progress in Physics, 78, 086901

\bibitem[\protect\citeauthoryear{Leclercq}{Leclercq}{2018}]{leclercq2018bayesian}
Leclercq F.,  2018, Physical Review D, 98, 063511

\bibitem[\protect\citeauthoryear{Lewis, Challinor  \& Lasenby}{Lewis
  et~al.}{2000}]{lewis2000efficient}
Lewis A.,  Challinor A.,   Lasenby A.,  2000, The Astrophysical Journal, 538,
  473

\bibitem[\protect\citeauthoryear{Loh}{Loh}{1996}]{loh1996latin}
Loh W.-L.,  1996, The annals of statistics, 24, 2058

\bibitem[\protect\citeauthoryear{Lopez-Paz \& Oquab}{Lopez-Paz \&
  Oquab}{2016}]{lopez2016revisiting}
Lopez-Paz D.,  Oquab M.,  2016, arXiv preprint arXiv:1610.06545

\bibitem[\protect\citeauthoryear{Lueckmann, Bassetto, Karaletsos  \&
  Macke}{Lueckmann et~al.}{2019}]{lueckmann2019likelihood}
Lueckmann J.-M.,  Bassetto G.,  Karaletsos T.,   Macke J.~H.,  2019, in
  Symposium on Advances in Approximate Bayesian Inference. pp 32--53

\bibitem[\protect\citeauthoryear{Mandelbaum}{Mandelbaum}{2018}]{mandelbaum2018weak}
Mandelbaum R.,  2018, Annual Review of Astronomy and Astrophysics, 56, 393

\bibitem[\protect\citeauthoryear{Marin, Pudlo, Robert  \& Ryder}{Marin
  et~al.}{2012}]{marin2012approximate}
Marin J.-M.,  Pudlo P.,  Robert C.~P.,   Ryder R.~J.,  2012, Statistics and
  Computing, 22, 1167

\bibitem[\protect\citeauthoryear{Mead, Peacock, Heymans, Joudaki  \&
  Heavens}{Mead et~al.}{2015}]{mead2015accurate}
Mead A.~J.,  Peacock J.~A.,  Heymans C.,  Joudaki S.,   Heavens A.~F.,  2015,
  Monthly Notices of the Royal Astronomical Society, 454, 1958

\bibitem[\protect\citeauthoryear{Miller, Cole, Forr{\'e}, Louppe  \&
  Weniger}{Miller et~al.}{2021}]{Miller2021truncated}
Miller B.~K.,  Cole A.,  Forr{\'e} P.,  Louppe G.,   Weniger C.,  2021,
  Advances in Neural Information Processing Systems, 34, 129

\bibitem[\protect\citeauthoryear{Nelder \& Mead}{Nelder \&
  Mead}{1965}]{nelder1965simplex}
Nelder J.~A.,  Mead R.,  1965, The computer journal, 7, 308

\bibitem[\protect\citeauthoryear{Papamakarios \& Murray}{Papamakarios \&
  Murray}{2016}]{papamakarios2016fast}
Papamakarios G.,  Murray I.,  2016, Advances in neural information processing
  systems, 29

\bibitem[\protect\citeauthoryear{Papamakarios, Pavlakou  \&
  Murray}{Papamakarios et~al.}{2017}]{papamakarios2017masked}
Papamakarios G.,  Pavlakou T.,   Murray I.,  2017, Advances in neural
  information processing systems, 30

\bibitem[\protect\citeauthoryear{Papamakarios, Sterratt  \&
  Murray}{Papamakarios et~al.}{2019}]{papamakarios2019sequential}
Papamakarios G.,  Sterratt D.,   Murray I.,  2019, in The 22nd International
  Conference on Artificial Intelligence and Statistics. pp 837--848

\bibitem[\protect\citeauthoryear{Park}{Park}{1994}]{park1994optimal}
Park J.-S.,  1994, Journal of statistical planning and inference, 39, 95

\bibitem[\protect\citeauthoryear{Porqueres, Heavens, Mortlock  \&
  Lavaux}{Porqueres et~al.}{2021a}]{porqueres2021bayesian}
Porqueres N.,  Heavens A.,  Mortlock D.,   Lavaux G.,  2021a, \mn@doi [Monthly
  Notices of the Royal Astronomical Society] {10.1093/mnras/stab204}, 502, 3035

\bibitem[\protect\citeauthoryear{Porqueres, Heavens, Mortlock  \&
  Lavaux}{Porqueres et~al.}{2021b}]{porqueres2021lifting}
Porqueres N.,  Heavens A.,  Mortlock D.,   Lavaux G.,  2021b, \mn@doi [Monthly
  Notices of the Royal Astronomical Society] {10.1093/mnras/stab3234}, 509,
  3194

\bibitem[\protect\citeauthoryear{Prangle}{Prangle}{2017}]{prangle2017adapting}
Prangle D.,  2017, Bayesian Analysis, 12, 289

\bibitem[\protect\citeauthoryear{Pritchard, Seielstad, Perez-Lezaun  \&
  Feldman}{Pritchard et~al.}{1999}]{pritchard1999population}
Pritchard J.~K.,  Seielstad M.~T.,  Perez-Lezaun A.,   Feldman M.~W.,  1999,
  Molecular biology and evolution, 16, 1791

\bibitem[\protect\citeauthoryear{Ribli, Pataki, Zorrilla~Matilla, Hsu, Haiman
  \& Csabai}{Ribli et~al.}{2019}]{ribli2019weak}
Ribli D.,  Pataki B.~{\'A}.,  Zorrilla~Matilla J.~M.,  Hsu D.,  Haiman Z.,
  Csabai I.,  2019, Monthly Notices of the Royal Astronomical Society, 490,
  1843

\bibitem[\protect\citeauthoryear{Rubin}{Rubin}{1984}]{rubin1984bayesianly}
Rubin D.~B.,  1984, The Annals of Statistics, pp 1151--1172

\bibitem[\protect\citeauthoryear{Schneider \& Hartlap}{Schneider \&
  Hartlap}{2009}]{schneider2009constrained}
Schneider P.,  Hartlap J.,  2009, Astronomy \& Astrophysics, 504, 705

\bibitem[\protect\citeauthoryear{Secco et~al.,}{Secco
  et~al.}{2022}]{secco2022dark}
Secco L.,  et~al., 2022, Physical Review D, 105, 023515

\bibitem[\protect\citeauthoryear{Sellentin \& Heavens}{Sellentin \&
  Heavens}{2018}]{sellentin2018insufficiency}
Sellentin E.,  Heavens A.~F.,  2018, Monthly Notices of the Royal Astronomical
  Society, 473, 2355

\bibitem[\protect\citeauthoryear{Sellentin, Heymans  \&
  Harnois-D{\'e}raps}{Sellentin et~al.}{2018}]{sellentin2018skewed}
Sellentin E.,  Heymans C.,   Harnois-D{\'e}raps J.,  2018, Monthly Notices of
  the Royal Astronomical Society, 477, 4879

\bibitem[\protect\citeauthoryear{Smyth \& Wolpert}{Smyth \&
  Wolpert}{1998}]{smyth1998evaluation}
Smyth P.,  Wolpert D.~H.,  1998, An evaluation of linearly combining density
  estimators via stacking.
Information and Computer Science, University of California, Irvine

\bibitem[\protect\citeauthoryear{Smyth \& Wolpert}{Smyth \&
  Wolpert}{1999}]{smyth1999linearly}
Smyth P.,  Wolpert D.,  1999, Machine Learning, 36, 59

\bibitem[\protect\citeauthoryear{Stein}{Stein}{1987}]{stein1987large}
Stein M.,  1987, Technometrics, 29, 143

\bibitem[\protect\citeauthoryear{Sugiyama et~al.,}{Sugiyama
  et~al.}{2022}]{sugiyama2022hsc}
Sugiyama S.,  et~al., 2022, Physical Review D, 105, 123537

\bibitem[\protect\citeauthoryear{Taylor, Joachimi  \& Kitching}{Taylor
  et~al.}{2013}]{taylor2013putting}
Taylor A.,  Joachimi B.,   Kitching T.,  2013, Monthly Notices of the Royal
  Astronomical Society, 432, 1928

\bibitem[\protect\citeauthoryear{Taylor, Kitching, Alsing, Wandelt, Feeney  \&
  McEwen}{Taylor et~al.}{2019}]{taylor2019cosmic}
Taylor P.~L.,  Kitching T.~D.,  Alsing J.,  Wandelt B.~D.,  Feeney S.~M.,
  McEwen J.~D.,  2019, Physical Review D, 100, 023519

\bibitem[\protect\citeauthoryear{Tegmark, Taylor  \& Heavens}{Tegmark
  et~al.}{1997}]{tegmark1997karhunen}
Tegmark M.,  Taylor A.~N.,   Heavens A.~F.,  1997, The Astrophysical Journal,
  480, 22

\bibitem[\protect\citeauthoryear{Upham, Brown  \& Whittaker}{Upham
  et~al.}{2021}]{upham2021sufficiency}
Upham R.~E.,  Brown M.~L.,   Whittaker L.,  2021, Monthly Notices of the Royal
  Astronomical Society, 503, 1999

\bibitem[\protect\citeauthoryear{Uria, C{\^o}t{\'e}, Gregor, Murray  \&
  Larochelle}{Uria et~al.}{2016}]{uria2016neural}
Uria B.,  C{\^o}t{\'e} M.-A.,  Gregor K.,  Murray I.,   Larochelle H.,  2016,
  The Journal of Machine Learning Research, 17, 7184

\bibitem[\protect\citeauthoryear{Wishart}{Wishart}{1928}]{wishart1928generalised}
Wishart J.,  1928, Biometrika, pp 32--52

\bibitem[\protect\citeauthoryear{Zuntz et~al.,}{Zuntz
  et~al.}{2015}]{zuntz2015cosmosis}
Zuntz J.,  et~al., 2015, Astronomy and Computing, 12, 45

\makeatother
\end{thebibliography}

%\begin{thebibliography}{99}
%\end{thebibliography}

%%%%%%%%%%%%%%%%%%%%%%%%%%%%%%%%%%%%%%%%%%%%%%%%%%

%%%%%%%%%%%%%%%%% APPENDICES %%%%%%%%%%%%%%%%%%%%%

%%%%%%%%%%%%%%%%%%%%%%%%%%%%%%%%%%%%%%%%%%%%%%%%%%

% Don't change these lines
\bsp	% typesetting comment
\label{lastpage}
\end{document}